\documentclass[3p,times,review]{elsarticle}

\usepackage{amssymb}

\usepackage[figuresright]{rotating}

\def\mb#1{\mbox {\boldmath {$#1$}}} 

\usepackage{stmaryrd,bm,amsmath}
\usepackage{dsfont}
\usepackage{amsbsy,amsmath,amsthm,latexsym,amssymb,mathtools}
\usepackage{graphicx,stmaryrd,sectsty}
\usepackage{setspace}
\usepackage[font=small,labelfont=bf]{caption}
\usepackage{epstopdf}
\usepackage{verbatim}
\usepackage{graphicx}
\usepackage{caption}
\usepackage{subcaption}
\usepackage{float}
\usepackage[toc,page]{appendix}
\usepackage{tikz}
\usepackage{enumitem}
\usepackage[utf8]{inputenc}

\newcommand{\B}{{Ba\v zant}}
\newcommand{\bc}{\begin{center}}
\newcommand{\ec}{\end{center}}
\newcommand{\bfr}{\begin{flushright}}
\newcommand{\efr}{\end{flushright}}

\newcommand{\be}{\begin{enumerate}}
\newcommand{\ee}{\end{enumerate}}
\newcommand{\bi}{\begin{itemize}}
\newcommand{\ei}{\end{itemize}}
\newcommand{\bd}{\begin{description}}
\newcommand{\ed}{\end{description}}
\newcommand{\beq}{\begin{equation}}
\newcommand{\eeq}{\end{equation}}
\newcommand{\bea}{\begin{eqnarray}}
\newcommand{\eea}{\end{eqnarray}}

\newcommand{\bfi}{\begin{figure}}
\newcommand{\efi}{\end{figure}}
\newcommand{\bay}{\begin{array}{l}}
\newcommand{\eay}{\end{array}}

\def\mb#1{\mbox {\boldmath {$#1$}}} 

\newcommand{\mbf}{\mathbf}

\newcommand{\cref}[1]{(\ref{#1})}   

\begin{document}

\begin{titlepage}
\clearpage\thispagestyle{empty}
\noindent
\hrulefill
\begin{figure}[h!]
\centering
\includegraphics[width=2 in]{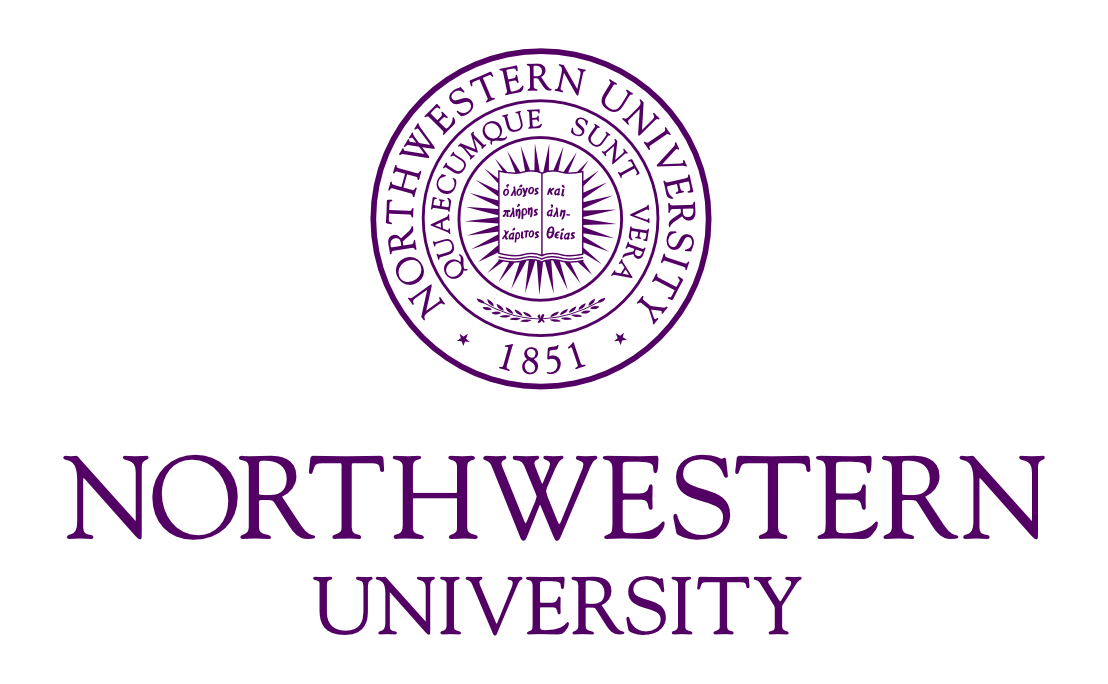}
\end{figure}
\begin{center}
{
{\bf Center for Sustainable Engineering of Geological and
Infrastructure Materials (SEGIM)} \\ [0.1in]
Department of Civil and Environmental Engineering \\ [0.1in]
McCormick School of Engineering and Applied Science \\ [0.1in]
Evanston, Illinois 60208, USA
}
\end{center} 
\hrulefill \\ \vskip 2mm
\vskip 0.5in
\begin{center}
{\large {\bf \uppercase{Adaptive Multiscale Homogenization of the Lattice Discrete Particle Model for the Analysis of Damage and Fracture in Concrete}}}\\[0.5in]
{\large {\sc Roozbeh Rezakhani, Xinwei Zhou, Gianluca Cusatis}}\\[0.75in]
{\sf \bf SEGIM INTERNAL REPORT No. 17-1/587A}\\[0.75in]
\end{center}
\noindent {\footnotesize {{\em Submitted to the International Journal of Solids and Structures \hfill January 2017} }}
\end{titlepage}

\begin{frontmatter}




\title{Adaptive Multiscale Homogenization of the Lattice Discrete Particle Model for the Analysis of Damage and Fracture in Concrete}


\author[l1]{Roozbeh Rezakhani}
\author[l2]{Xinwei Zhou}
\author[l1]{Gianluca Cusatis \footnote{Corresponding author. \\ E-mail address: g-cusatis@northwestern.edu}}
\address[l1]{Department of Civil and Environmental Engineering, Northwestern University, Evanston, IL 60208, USA.}
\address[l2]{ES3, 550 West C St., San Diego, CA 92101, U.S.A}

\begin{abstract}
This paper presents a new adaptive multiscale homogenization scheme for the simulation of damage and fracture in concrete structures. A two-scale homogenization method, coupling meso-scale discrete particle models to macro-scale finite element models, is formulated into an adaptive framework. A continuum multiaxial failure criterion for concrete is calibrated on the basis of fine-scale simulations, and it serves as the adaptive criterion in the multiscale framework. Thus, in this approach, simulations start without assigning any material Representative Volume Element (RVE) to the macro-scale finite elements. The finite elements that meet the adaptive criterion and must be entered into the multiscale homogenization framework are detected on the fly. This leads to a substantial reduction of the computational cost especially for loading conditions leading to damage localization in which only a small portion of the FE mesh is enriched with the homogenized RVE. Several numerical simulations are carried out to investigate the capability of the developed adaptive homogenization method. In addition, a detailed study on the computational cost is performed.
\end{abstract}


\end{frontmatter}

\section{Introduction}
All natural and man-made engineering materials are heterogeneous at a certain length scale. Macroscopic mechanical properties of materials, such as Young's modulus, tensile and compressive strengths, hygrothermal characteristics, etc., strongly depend on the features of their constitutive heterogeneities. For instance, mechanical properties of concrete, the most used engineering material on earth, directly depend on the mechanical properties of aggregate, cement, admixtures, aggregate size distribution, and aggregate volume fraction. Therefore, employing detailed fine-scale models which take material heterogeneities into account is crucial in order to obtain accurate and informative numerical results. However, the numerical simulation of entire large structural systems such as dams, bridge piers, and nuclear power plants by fine-scale models lead to an enormous number of degrees of freedom and a prohibitive computational cost. This has motivated researchers to develop different categories of multiscale computational frameworks, by which one can benefit from utilizing an accurate mechanical model in the analysis of a large engineering problem with a reasonable computational cost. Multiscale analysis frameworks fall into two primary groups \cite{Unger}. The first group includes hierarchical multiscale models, in which a fine-scale problem, simulated by a continuous or discrete mechanical model, is coupled to a macro-scale problem generally simulated by FEM. Hierarchy of the computational framework is established through solving the macro- and fine-scale problems independently, while information flows between the scales during the analysis. The second category includes concurrent multiscale models, in which the problem domain is decomposed into two subdomains. One subdomain is treated as homogeneous continuum modeled by FEM over which deformations are considered to be in the elastic range, while the other subdomain is modeled by the detailed fine-scale model of interest. These two subdomains are connected through appropriate constraints, and the two scales are solved simultaneously.

Multiscale homogenization is a hierarchical method which has been intensively investigated over the past decades. Analytical homogenization methods were first developed to derive mathematical equations for the effective material properties of heterogeneous solids based on the exact solution of the fine-scale boundary value problems \cite{Hill-1,Eshelby-1,Hashin-1,Christensen-1}. However, these approaches are limited to elastic materials with simple periodic internal structures studied under small deformation fields. Computational homogenization methods were then developed to conquer these limitations. In these multiscale frameworks, a Representative Volume Element (RVE) of the fine-scale material is constructed and assigned to each macroscopic computational point. The RVE homogenized response is considered as the effective material behavior, and no specific constitutive equations are defined at the macro-scale. During each computational step, the macroscopic strain tensor at a FE Gauss point is applied to the related material RVE as boundary condition. Next, the RVE solution is used to calculate the homogenized stress tensor, which is then transferred back to the macroscopic computational point to continue the analysis. Computational homogenization has been widely used in the analysis of complicated engineering problems \cite{Smith-1,Feyel-1,Kouznetsova-1,Miehe-1}. Smith et al.  \cite{Smith-1} developed a multilevel finite element method, in which both RVE and macroscopic domains are meshed with FE (FE$^2$), and employed it in the analysis of perforated viscoelastic materials. Feyel \cite{Feyel-1} established an FE$^2$ multiscale homogenization framework for generalized continua, in which curvature and moment stress tensors are taken into account . Kouznetsova \cite{Kouznetsova-1} formulated a higher order homogenization scheme in which first and second order gradients of the macroscopic deformation field are applied on the RVE as boundary conditions. It is shown that in the gradient enhanced homogenization method, non-uniform macroscopic deformation fields can be reproduced within the RVE boundary value problem. Miehe et al. \cite{Miehe-1} used the computational homogenization in the analysis of polycrystalline materials under large plastic deformations. In addition, computational homogenization method has been used to couple discrete element models (DEM) to FE. Guo et al. \cite{Guo-1} studied nonlinear and dissipative response of granular media under monotonic and cyclic loading condition by a coupled FEM/DEM approach. Wang \cite{Wang-1} recently developed a hierarchical multiscale hygro-mechanical model to investigate fluid flow in granular media.

Asymptotic Expansion Homogenization (AEH), a more mathematically thorough approach, employs the asymptotic expansion of the field variables to derive the governing equilibrium equations and the corresponding boundary conditions for the macro- and the RVE-scale boundary value problems \cite{Chung-1}. Fish et al. \cite{Fish-3} utilized this approach to study the behavior of atomistic systems using scale separation in both space and time. AEH theory is easy to implement in computational softwares and has been extensively used by researchers to study the mechanical response of materials. Hassani and Hinton \cite{Hassani-1, Hassani-2, Hassani-3} published a three-paper review, in which they presented details of AEH theory and its finite element implementation for elastic materials with periodic microstructure. The presented theory was then used in composite material as well as structural topology optimization. Fish et al. formulated an AEH framework to investigate elasto-plastic behavior of composites \cite{Fish-2}. Caglar and Fish \cite{Fish-1} proposed a reduced order AEH approach to simulate nonlinear behavior of fiber composite materials with decreased computational cost. Ghosh et al. combined the asymptotic expansion approach with Voronoi Cell Finite Element Method (VCFEM) and developed a multiscale theoretical framework to approximate elastic properties \cite{Ghosh-1} and investigate elasto-plastic response \cite{Ghosh-2} of composites with random meso-structure. 

Within the current literature on the multiscale homogenization methods, computational accuracy of the developed framework has been more investigated and praised compared to the computational efficiency. One of the major shortcomings of the classical homogenization methods, in which material RVEs are assigned to all macro-scale FEs from the beginning of the analysis, is the tremendous computational expenses necessary to solve the boundary value problems at each integration point at separate scales during the analysis. In engineering problems dealing with fracture and failure in concrete structures, cracking and damage usually localizes in a certain region of the structure, and the rest of the material domain remains elastic. Therefore, assigning material RVEs to all macroscopic integration points is superfluous. In this regard, Ghosh et al. \cite{Ghosh-3,Ghosh-4} proposed an adaptive concurrent multi-level model for multiscale simulation of fracture in heterogeneous composites, in which three modeling resolution levels are used. The macroscopic model uses continuum damage mechanics, the RVE model is based on asymptotic homogenization, and the microscale simulations of composites are performed by VCFEM. The framework is formulated on the basis of  energy criteria governing the transition between adjacent scales. Regarding analysis of concrete structures, Sun and Li \cite{Sun-1} developed an adaptive concurrent multiscale FEM in which any region of the macroscopic FE model that meets a specific criteria is replaced by a fine-scale discretization accounting for heterogeneity. In this multiscale model, concrete behavior is considered to be elastic up to the peak and the adaptive framework is limited to simple 2D problems.

The present study extends the discrete to continuum homogenization method recently developed by the authors \cite{Rezakhani-JMPS} to an adaptive framework specialized for concrete. In \cite{Rezakhani-JMPS}, Rezakhani and Cusatis presented a homogenization framework to couple the Lattice Discrete Particle Model (LDPM) \cite{cusatis-ldpm-1,cusatis-ldpm-2} to FE, in which material RVEs simulated by LDPM are assigned to all macroscopic FEs ahead of the analysis. Two-scale numerical examples were solved, and computational accuracy of the developed multiscale framework was verified by comparing the results with the full fine-scale simulations. In the current paper, all macroscopic FEs are initialized with an isotropic linear elastic constitutive equation. During the simulation, an adaptive scheme automatically detects the FEs that meet a certain criteria and need to be assigned a material RVE. This strategy leads to a considerable saving in the computational cost while the accuracy is still demonstrated to be satisfactory.

\section{Review of the Lattice Discrete Particle Model (LDPM) for concrete}	
\begin{figure}[t!]
\centering 
{\includegraphics[width=\textwidth]{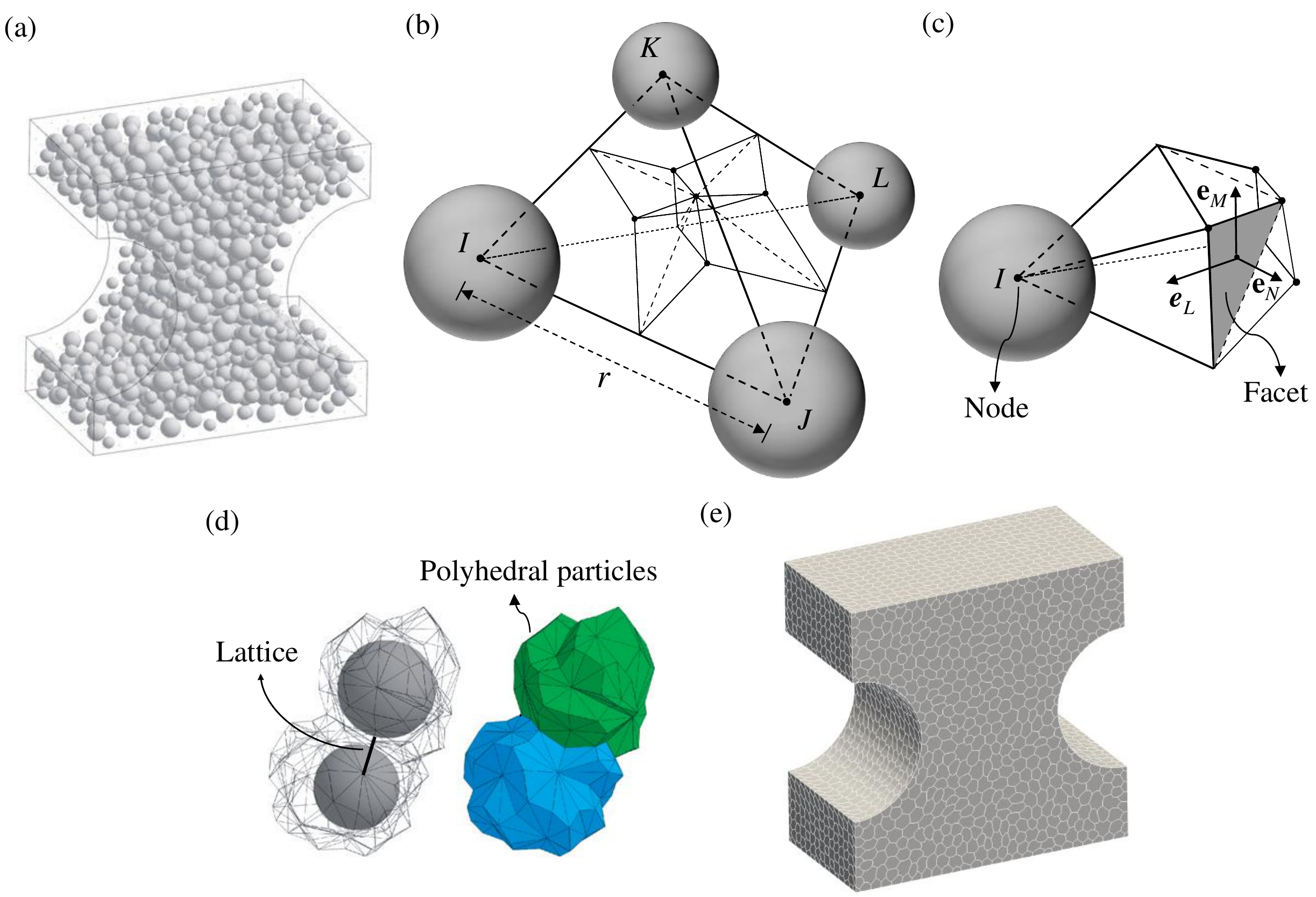}}
\caption{(a) Spherical aggregate distribution in a typical dogbone specimen. (b) A typical LDPM tetrahedron connecting four adjacent aggregates and its associated tessellation. (c) Tetrahedron portion associated with aggregate $I$. (d) LDPM polyhedral particle enclosing spherical aggregate pieces. (e) Polyhedral particle representation of a typical dogbone specimen.}
\label{LDPM}
\end{figure}

The Lattice Discrete Particle Model (LDPM) constructs the geometrical representation of concrete meso-structure through the following steps: (1) Coarse aggregate pieces, whose shapes are assumed to be spherical, are introduced into the concrete volume by a try-and-reject random procedure. Aggregate diameters are determined by sampling an assumed aggregate size distribution function. Spherical aggregates distribution in a typical dogbone specimen is depicted in Figure \ref{LDPM}a. (2) Zero-radius aggregate pieces (nodes) are randomly distributed over the external surfaces to facilitate the application of boundary conditions. (3) Delaunay tetrahedralization of the generated aggregate centers and the associated three-dimensional domain tessellation are then carried out to obtain a network of triangular facets inside each tetrahedral element as shown in Figure \ref{LDPM}b. Figure \ref{LDPM}c illustrates a portion of the tetrahedral element associated with one of its four nodes $I$ and the corresponding facets. Combining such portions from all tetrahedral elements connected to the same node $I$, one obtains the corresponding polyhedral particle which encloses the spherical aggregate. Two adjacent polyhedral particles interacting through shared triangular facets are depicted in Figure \ref{LDPM}d. Polyhedral particles are considered to be rigid,  and triangular facets, on which strain and stress quantities are defined in vectorial form, are assumed to be the potential material failure locations. Figure \ref{LDPM}e presents the polyhedral particle representation of the typical dogbone specimen. One should consider that spherical aggregates are generated to build a discrete model which resembles concrete real meso-structure, while they are not directly used in the numerical solution procedure. Centroid of the spherical aggregates, called ``node'' for the rest of this paper, and associated polyhedral particles are the geometrical units that are employed in the numerical analysis. Three sets of equations are necessary to complete the discrete model framework: definition of strain on each facet, constitutive equation which relates facet stress vector to facet strain vector, and particle equilibrium equations.
\newline \\
\noindent \textbf{Facet strain definition.} Rigid body kinematics is employed to describe the deformation of the lattice/particle system, and the displacement jump, $\llbracket \mathbf{u}_{C} \rrbracket$, at the centroid of each facet is used to define measures of strain as 
\begin{equation} \label{eps} 
\epsilon_{\alpha}=\frac{1}{r} \llbracket \mathbf{u}_{C} \rrbracket \cdot \mathbf{e}_{\alpha}
\end{equation}
where $\alpha=N,M,L$ ($\epsilon_N=$ facet normal strain component;  $\epsilon_M$ and $\epsilon_L=$ facet tangential strain components); $r=$ length of the line that connects the nodes sharing the facet and also the associated tetrahedron edge (see Figure \ref{LDPM}b) ; $\mathbf{e}_{\alpha}$ are unit vectors defining a facet local Cartesian system of reference such that $\mathbf{e}_{N}=$ is orthogonal to the facet, and $\mathbf{e}_{M}$ and $\mathbf{e}_{L}=$ are the facet tangential unit vectors (see Figure \ref{LDPM}c).
\newline \\
\noindent \textbf{Facet vectorial constitutive equations.} Next, a vectorial constitutive law governing the behavior of the material is imposed at the centroid of each facet. In the elastic regime, the normal and shear stresses are proportional to the corresponding strains: $t_{N} = E_N \epsilon_{N}; ~  t_{M} = E_T \epsilon_{M}; ~  t_{L} = E_T \epsilon_{L}$, where $E_N=E_0$; $E_T=\alpha E_0$; $E_0=$ effective normal modulus; $\alpha=$ shear-normal coupling parameter. For stresses and strains beyond the elastic limit, concrete mesoscale nonlinear phenomena are characterized by three mechanisms and the corresponding facet level vectorial constitutive equations are briefly described below. 

\begin{itemize}[leftmargin=*]
\item{\textit{Fracture and cohesion due to tension and tension-shear.} For tensile loading ($\epsilon_N>0$), the fracturing behavior is formulated through an effective strain, $\epsilon = \sqrt{\epsilon_N^{2}+\alpha (\epsilon_M^{2} + \epsilon_L^{2})}$, and stress, $t = \sqrt{{ t _{N}^2+  (t _{M}^2+t _{L}^2) / \alpha}}$, which are used to define the facet normal and shear stresses as \mbox{$t _{N}= \epsilon_N(t / \epsilon)$}; \mbox{$t _{M}=\alpha \epsilon_{M}(t / \epsilon)$}; \mbox{$t _{L}=\alpha \epsilon_{L}(t / \epsilon)$}. The effective stress $t$ is incrementally elastic ($\dot{t}=E_0\dot{\epsilon}$) and must satisfy the inequality $0\leq t \leq \sigma _{bt} (\epsilon, \omega) $ where 

\begin{equation}
\sigma_{bt} = \sigma_0(\omega) \exp \left[-H_0(\omega)  \langle \epsilon_{max}-\epsilon_0(\omega) \rangle / \sigma_0(\omega)\right]
\end{equation}
in which $\langle x \rangle=\max \{x,0\}$; $\epsilon_0(\omega) = \sigma_0(\omega)/E_0$; $\tan(\omega) =\epsilon _N / \sqrt{\alpha} \epsilon_{T}$ = $t_N \sqrt{\alpha} / t_{T}$ in which $\epsilon_T=\sqrt{\epsilon_M^{2} + \epsilon_L^{2}}$ and $t_T=\sqrt{t_M^{2} + t_L^{2}}$.  $\omega$ is the parameter that defines the degree of interaction between shear and normal loading. $\epsilon_{max} = \sqrt{\epsilon_{N,max}^{2}+\alpha \epsilon_{T,max}^{2}}$ is a history dependent variable, and $\epsilon_{max} = \epsilon$ in the absence of unloading. The post peak softening modulus is defined as $H_{0}(\omega)=H_{t}(2\omega/\pi)^{n_{t}}$, where $n_t$ is the softening exponent, $H_{t}$ is the softening modulus in pure tension ($\omega=\pi/2$) expressed as $H_{t}=2E_0/\left(l_t/r-1\right)$; $l_t=2E_0G_t/\sigma_t^2$; and $G_t$ is the mesoscale fracture energy. LDPM provides a smooth transition between pure tension and pure shear ($\omega=0$), with a parabolic variation for strength given by (solid curve in Figure \ref{LDPM-constitutive}a)

\begin{equation}
\sigma_{0}(\omega )=\sigma _{t}r_{st}^2\Big(-\sin(\omega) + \; \sqrt{\sin^2(\omega)+4 \alpha \cos^2(\omega) / r_{st}^2}\Big) \; / \; [2 \alpha \cos^2(\omega)]
\end{equation}
where $r_{st} = \sigma_s/\sigma_t$ is the shear to tensile strength ratio. The dashed line in Figure \ref{LDPM-constitutive}a represents the strength domain $\sigma_{bt}$ corresponding to $\epsilon_{max} = 4\sigma_t/E_0$. Facet $t_N$ and $t_T$ versus $\epsilon_N$ and $\epsilon_T$ are shown in Figure \ref{LDPM-constitutive}b for pure tension ($\omega = 0$), pure shear ($\omega = \pi/2$), and $\omega = \pi/8$ using $r$=10 mm, $\alpha$=0.25, $E_0$=30 GPa, $\sigma_t$=3 MPa, $\sigma_s$=4.5 MPa, $l_t$=100 mm, and $n_t$=0.2. Finally, Figure \ref{LDPM-constitutive}c represents the unloading-reloading rule adopted in this work in terms of the effective stress versus the effective strain relationship in which $\epsilon_{tr} = k_t (\epsilon_{max} - \sigma_{bt}/E_0)$, and $k_t$ is a material parameter that defines the size of the hysteresis cycle.}

\begin{figure}[t!]
\centering 
{\includegraphics[width=\textwidth]{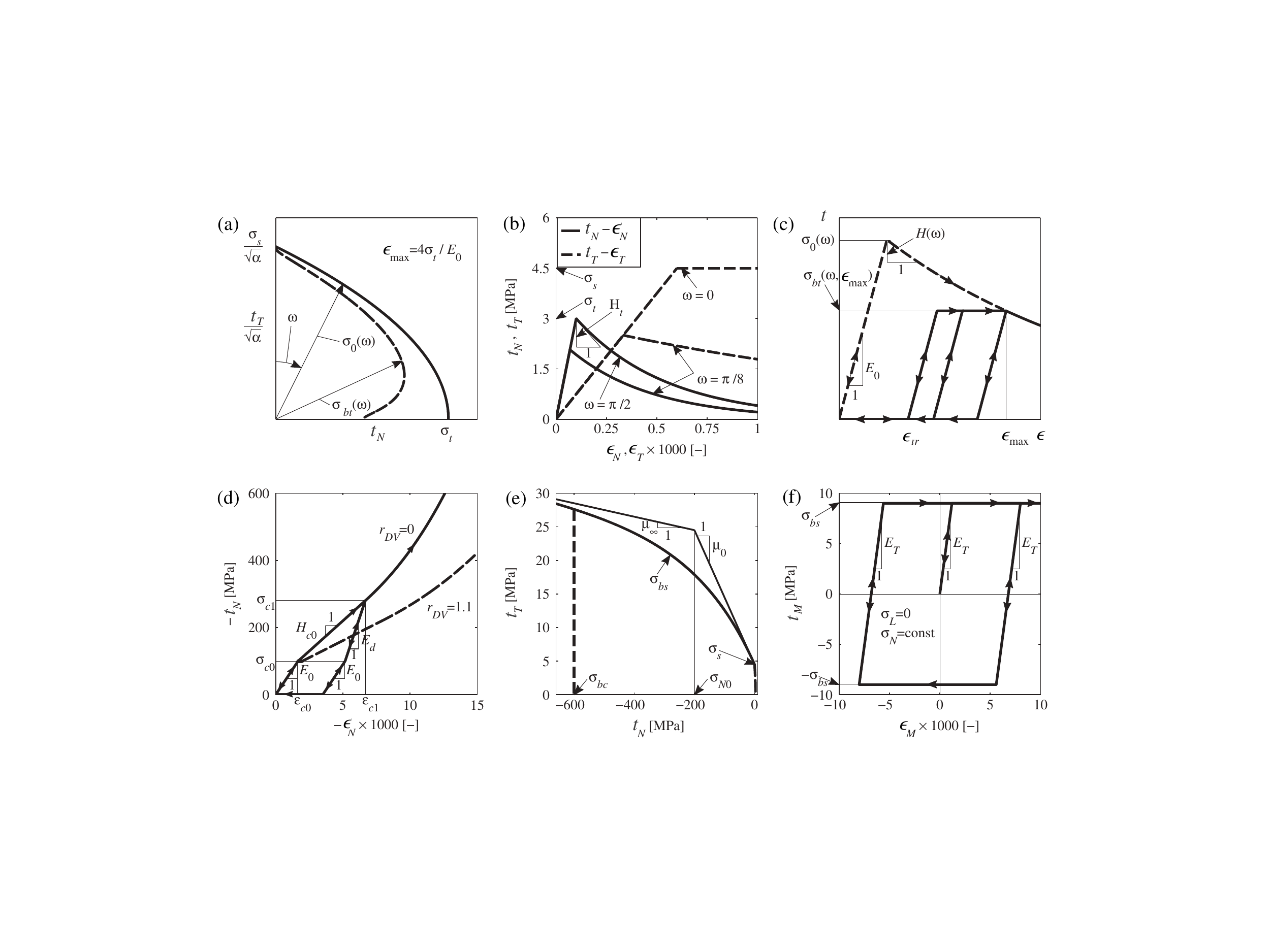}}
\caption{LDPM facet constitutive laws: (a) Initial and damaged shear strength as function of positive normal stress. (b) Typical stress versus strain curves on an LDPM facet. (c) Unloading-reloading model. (d) Typical compressive normal stress versus normal strain curves. (e) Shear strength as a function of compressive stresses (frictional behavior). (f) Typical shear stress versus shear strain curve.}
\label{LDPM-constitutive}
\end{figure}

\item{\textit{Compaction and pore collapse in compression.} Normal stresses for compressive loading ($\epsilon_N<0$) are computed through the inequality $-\sigma_{bc}(\epsilon_D, \epsilon_V)\leq t_N \leq 0$, where $\sigma_{bc}$ is a strain-dependent boundary that depends on the volumetric strain, $\epsilon_V$, and the facet deviatoric strain, $\epsilon_D=\epsilon_N-\epsilon_V$. The volumetric strain is computed by the volume variation of the tetrahedral element as $\epsilon_V= \Delta V/ 3V_0$ and is assumed to be constant for all facets belonging to a given tetrahedron. Beyond the elastic limit, $\sigma_{bc}$ models pore collapse for $(\epsilon_{c0} \leq -\epsilon_V \leq \epsilon_{c1})$ as a linear evolution of stress for increasing volumetric strain with stiffness $H_{c}$ and compaction and rehardening beyond the pore collapse limit for ($-\epsilon_V \geq \epsilon_{c1}$). $\epsilon_{c0} = \sigma_{c0}/E_0$ is the compaction strain at which pore collapse starts, and $\epsilon_{c1} = \kappa_{c0} \epsilon_{c0}$ is the compaction strain at the beginning of rehardening. $\kappa_{c0}$ is a material parameter, and $\sigma_{c0}$ is the mesoscale compressive yield stress. Therefore, one can write
\begin{equation}
\sigma_{bc}(\epsilon_D,\epsilon_V) = 
\begin{cases}
    \sigma_{c0} & \text{for } (-\epsilon_V < 0) \\
    \sigma_{c0} + \langle -\epsilon_V-\epsilon_{c0}\rangle H_c(r_{DV}) & \text{for }  (0 \leq-\epsilon_V \leq \epsilon_{c1}) \\
    \sigma_{c1}(r_{DV}) \exp \left[( -\epsilon_{V}-\epsilon_{c1} ) H_c(r_{DV})/\sigma_{c1}(r_{DV}) \right] & \text{otherwise}
\end{cases}
\end{equation}
where $H_c=(H_{c0}-H_{c1})/(1 + \kappa_{c2} \left\langle r_{DV} - \kappa_{c1} \right\rangle)+H_{c1}$; $r_{DV}=\vert\epsilon_D\vert/\epsilon_{V0}$ for ($\epsilon_{V}>0$) and $r_{DV}=-\vert\epsilon_D\vert/(\epsilon_V-\epsilon_{V0})$ for ($\epsilon_{V}\leq0$), in which $\epsilon_{V0}=\kappa_{c3}\epsilon_{c0}$. $H_{c0}$, $H_{c1}$, $\kappa_{c1}$, $\kappa_{c2}$, $\kappa_{c3}$ are material parameters. These boundaries are shown in Figure \ref{LDPM-constitutive}d as the compressive normal stress versus strain for $r_{DV} = 0$ and $r_{DV} = 1.1$. The unloading-reloading path is also shown for the case of $r_{DV} = 0$ in which $E_d$ is the densified normal modulus and is a material property. LDPM parameters considered in plotting these curves are $E_0$= 60 GPa, $\sigma_{c0}$=100 MPa, $H_{c0}/E_0$= 0.6, $H_{c1}/E_0$= 0.1, $k_{c0}$= 4, $E_d/E_0$= 2, $k_{c1}$= 1, $k_{c1}$= 0.5, and $k_{c3}$= 0.1.}

\item{\textit{Friction due to compression-shear.} The incremental shear stresses in presence of compression are computed as  $\dot{t}_M=E_T(\dot{\epsilon}_M-\dot{\epsilon}^{p}_M)$ and \mbox{$\dot{t}_L=E_T(\dot{\epsilon}_L-\dot{\epsilon}^{p}_L)$}, where  \mbox{$\dot{\epsilon}_M^{p}=\dot{\lambda} \partial \varphi / \partial t_M$}, \mbox{$\dot{\epsilon}_L^{p}=\dot{\lambda} \partial \varphi / \partial t_L$}, and $\lambda$ is the plastic multiplier with loading-unloading conditions  $\varphi \dot{\lambda} \leq 0$ and $\dot{\lambda} \geq 0$. The plastic potential is defined as \mbox{$\varphi=\sqrt{t_M^2+t_L^2} - \sigma_{bs}(t_N)$}, where the nonlinear frictional law for the shear strength is assumed to be 

\begin{equation}
\sigma_{bs} = \sigma_s + (\mu_0 - \mu_\infty)\sigma_{N0}[1 - \exp(t_N / \sigma_{N0})] - \mu_\infty t_N
\end{equation}
where $\sigma_{N0}$ is the transitional normal stress; $\mu_0$ and $\mu_\infty$ are the initial and final internal friction coefficients. Shear strength envelope $\sigma_{bs}$ and typical shear stress versus strain relationship are shown in Figures \ref{LDPM-constitutive}e and \ref{LDPM-constitutive}f, respectively.}
\end{itemize}

\noindent \textbf{Particle equilibrium equations.} Finally, the governing equations of the LDPM framework are completed through the equilibrium equations of each individual particle. Linear and angular momentum balance equations for a generic polyhedral particle can be written as
\begin{equation} \label{motion-1}
\sum_{\mathcal{F}} A \mathbf{t} + V \mathbf{b}^0 = 0; \hspace{0.5 in} \sum_{\mathcal{F}} A \mathbf{w} = 0
\end{equation}
where $\mathcal{F}$ is the set of facets that form the polyhedral particle; $A$ = facet area; superimposed dots represent time derivatives; $V$ is the particle volume;  $\mathbf{b}^0$ is the body force vector; $\mathbf{t} = t_{\alpha}\mathbf{e}_{\alpha} = t_N \mathbf{e}_{N} + t_M \mathbf{e}_{M} + t_L \mathbf{e}_{L}$ is the resultant traction vector applied on each triangular facet; $\mathbf{w}$ is the moment of $\mathbf{t}$ with respect to the node which is enclosed by the polyhedral particle.

LDPM is implemented in a computational software named MARS \cite{mars-1} and has been used successfully to simulate concrete behavior in different types of laboratory experiments \cite{cusatis-ldpm-2}. Furthermore, LDPM has shown superior capabilities in modeling concrete behavior under dynamic loading \cite{cusatis-Jovanca}, Alkali Silica Reaction (ASR) deterioration \cite{cusatis-mohammed}, fracture simulation of FRP reinforced concrete \cite{Chiara}, as well as failure and fracture of fiber-reinforced concrete \cite{cusatis-Ed1,cusatis-Ed2}. 

\section{Multiscale homogenization method}	
Multiscale homogenization theories are based on two primary assumptions: (1) A certain volume of the material, whose effective mechanical behavior is completely representative of the material response,  can be identified. In homogenization theories, this volume of material is  called the Representative Volume Element (RVE), in which the internal features of the material structure are modeled explicitly \cite{Kouznetsova-1}. (2) The ``separation of scales'' hypothesis is valid, which means that the ratio of the RVE size to the characteristic size of the macroscopic problem is very small. Furthermore, the ratio of the RVE size to the characteristic length of the fine-scale material heterogeneity, e.g. maximum aggregate size in granular media, is large enough so that the assigned RVE is an appropriate representative of the underlying material. 

\begin{figure}[t!]
\centering 
{\includegraphics[width=\textwidth]{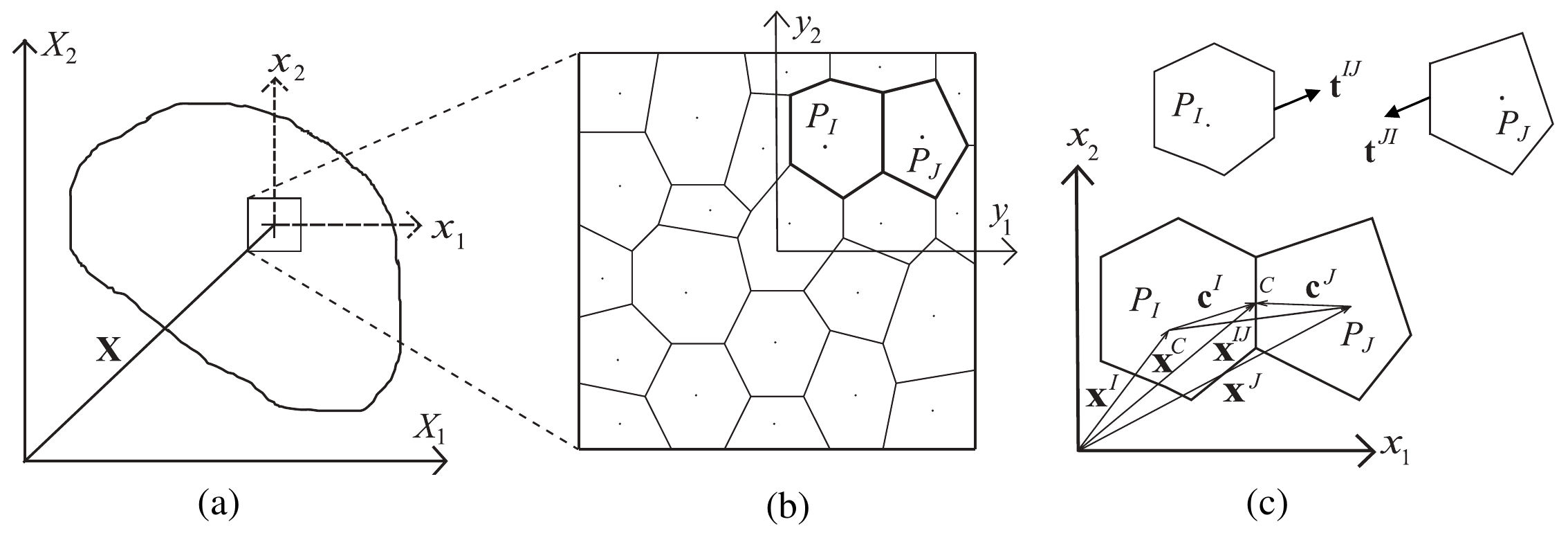}}
\caption{Geometrical explanation of the two-scale problem: (a) Macro material domain. (b) Mesoscale domain with material heterogeneity. (c) Geometry of two neighboring particles.}
\label{TwoScaleAnalysis}
\end{figure}

General definition of the two-scale homogenization problem is depicted in Figure \ref{TwoScaleAnalysis}. In Figure \ref{TwoScaleAnalysis}a, a generic macroscopic material domain is illustrated in a global macro-scale coordinate system, denoted by $\bf X$. Material is considered to be homogeneous at this scale, and no heterogeneity is taken into account. An enlarged view of the material structure at an arbitrary point in the macro-scale domain is depicted in Figure \ref{TwoScaleAnalysis}b as a representative volume of the heterogeneous material. One can see that the material heterogeneity is taken into account at this scale, the so-called fine-scale, and is modeled by means of a discrete meso-scale particle model. Therefore, two separate length scales and the corresponding local coordinate systems, $\bf x$ and  $\bf y$, are designated at any point of the material domain to serve as the local macro- and meso-scale problems, respectively. According to the separation of scales hypothesis, the macro- and meso-scale coordinate systems are linked as follows
\begin{eqnarray}\label{scale-link-1}
\mathbf{x}=\eta \mathbf{y}  \hspace{0.5 in}   0< \eta <<1
\end{eqnarray}
where $\eta$ is a very small positive scalar. Interaction of two neighboring particles, $I$ and $J$, which share a generic facet is shown in Figure \ref{TwoScaleAnalysis}c. The strain definition at each facet, Equation \ref{eps}, can be written in terms of the displacement and rotation vectors of the two particles sharing that facet. For the facet shared between particles $I$ and $J$ one can write
\begin{equation} \label{eps-1} 
\epsilon^{IJ}_{\alpha}=\frac{1}{r} \left(\mathbf{U}^J + \mb \Theta^{J} \times \mathbf{c}^{J} - \mathbf{U}^I - \mb{\Theta}^I \times \mathbf{c}^I \right) \cdot \mathbf{e}^{IJ}_{\alpha} 
\end{equation}
where $r=|\mathbf{x}^{IJ}|$; $\mathbf{x}^{IJ}=\mathbf{x}^J-\mathbf{x}^I$ is the vector connecting the particle nodes $P_I$ and $P_J$; $\mathbf{e}^{IJ}_{\alpha}$ are unit vectors defining Cartesian system of reference on facet $IJ$ such that $\mathbf{e}^{IJ}_{N}=$ is orthogonal to the facet and $\mathbf{e}^{IJ}_{N} \cdot \mathbf{x}^{IJ} >0$; $\mathbf{U}^I$, $\mathbf{U}^J$ = displacement vectors of node $P_I$ and $P_J$; $\mb{\Theta}^{I}$, $\mb{\Theta}^{J}$ = rotation vectors of node $P_I$ and $P_J$; and $\mathbf{c}^{I}$, $\mathbf{c}^{J}$ = vectors connecting nodes $P_I$ and $P_J$ to the facet centroid, see Fig. \ref{TwoScaleAnalysis}c. It must be noted here that displacements and rotations are assumed to be independent variables. In addition, Equation \ref{eps-1} is limited to the case of small strains and displacements, which is a rational assumption in absence of large plastic deformation prior to fracture as observed in brittle and quasi-brittle materials

\noindent\textbf{Asymptotic expansion homogenization}. The main goal in multiscale homogenization theory is deriving the governing equations that govern the problem at different length scales. To accomplish this goal, employing asymptotic form of the problem field variables is one of the most renowned methods. Therefore, for the current case of mechanical behavior of discrete particulate media, asymptotic expansion of the particle displacement and rotation fields is taken into account. Following the work of Rezakhani and Cusatis \cite{Rezakhani-JMPS}, displacement and rotation of a generic node P$_I$,  $\bold{U}^I = \bold{u}(\bold{x}^I, \bold{y}^I)$ and ${\boldsymbol \Theta}^I = \boldsymbol{\theta}(\bold{x}^I, \bold{y}^I)$, can be approximated by virtue of the following asymptotic expansions
\begin{equation}
\bold{u}(\bold x, \bold y) \approx \bold u^0(\bold x, \bold y)+\eta \bold u^1(\bold x, \bold y) 
\label{disp-expansion}
\end{equation}
\begin{equation}
\boldsymbol{\theta}  (\bold x, \bold y)  \approx \eta^{-1} \boldsymbol{\omega} ^{0}(\bold x, \bold y) +  \boldsymbol{\varphi}^0(\bold x, \bold y)+  \boldsymbol{\omega}^{1}(\bold x,\bold y)+\eta \boldsymbol{\varphi}^{1}(\bold x, \bold y) 
\label{rot-expansion}
\end{equation}

Terms of order $\mathcal{O}(\eta^2)$ and higher are neglected in the above expansions. $\bold  u^0(\bold x, \bold y)$, and $\bold  u^1(\bold x, \bold y)$ are coarse- and fine-scale displacement vectors, respectively, which are continuous functions with respect to $\bold{x}$ and discrete (i.e. defined only at finite number of points) with respect to $\bold{y}$. $\boldsymbol{\omega}^{0}$, $\boldsymbol{\omega}^{1}$ are rotations in the fine-scale coordinate system, whereas $\boldsymbol{\varphi}^{0}$ and $\boldsymbol{\varphi}^{1}$ are the corresponding coarse-scale rotations. Derivation of the asymptotic expansion of rotation is explained in \cite{Rezakhani-JMPS} in details.

The distance between nodes $P_I$ and $P_J$ can be considered infinitesimal in the macroscopic coordinate $\mathbf{x}$. Therefore, the displacement and rotation of the node $P_J$ at coordinate $\mathbf{x}^J$ can be written in the form of Taylor series expansion around the node $P_I$ at coordinate $\mathbf{x}^I$ in the macroscopic coordinate system $\bf x$. Accepting macroscopic continuity and differentiablity of the displacement and rotation vectors, one can write $U_i^J=u_i(\mathbf{x}^J,\mathbf{y}^J) = u_i^J+u^J_{i,j} \,x^{IJ}_j + \frac{1}{2}u^J_{i,jk}\, x^{IJ}_j x^{IJ}_k +\cdots$ and $\Theta_i^J=\theta_i(\mathbf{x}^J,\mathbf{y}^J)= \theta_i^J+\theta_{i,j}^J x^{IJ}_j +\frac{1}{2} \theta^J_{i,jk}\, x^{IJ}_j x^{IJ}_k +\cdots$ in which $u_i^J=u_i(\mathbf{x}^I,\mathbf{y}^J)$ and $\theta_i^J= \theta_i(\mathbf{x}^I,\mathbf{y}^J)$. By substituting these expansions and the Equations \ref{disp-expansion} and \ref{rot-expansion} into the expression of the facet $IJ$ strains, Equation \ref{eps-1}, and collecting terms of the same order, multiple scale definition of the facet strains is obtained as follows
\begin{equation}\label{eps-expansion}
\epsilon_{\alpha}=\eta^{-1} \epsilon_{\alpha}^{-1} + \epsilon_{\alpha}^0 + \eta \epsilon_{\alpha}^1
\end{equation}
where
\begin{equation}\label{eps-expansion-minus}
\epsilon_{\alpha}^{-1} = \bar{r}^{-1} \bigg[  u_i^{0J} - u^{0I}_i + \varepsilon_{ijk} \omega_j^{0J} \bar c_{k}^{J} - \varepsilon_{ijk} \omega_j^{0I} \bar c_{k}^{I} \bigg]e^{IJ}_{\alpha i} 
\end{equation}
\begin{equation}\label{eps-expansion-zero}
\epsilon_{\alpha}^0 =\bar{r}^{-1} \bigg[  u_i^{1J} + u^{0J}_{i,j} y^{IJ}_j - u^{1I}_i + \varepsilon_{ijk} \bigg( \varphi_j^{0J} + \omega_j^{1J} + \omega_{j,m}^{0J} y^{IJ}_m  \bigg) \bar c_{k}^{J} -  \varepsilon_{ijk} \bigg( \varphi_j^{0I} + \omega_j^{1I} \bigg) \bar c_{k}^{I} \bigg] e^{IJ}_{\alpha i} 
\end{equation}
\begin{equation}\label{eps-expansion-plus}
\begin{split}
\epsilon_{\alpha}^1 = \bar{r}^{-1} \bigg[  u^{1J}_{i,j} y^{IJ}_j + \frac{1}{2}u^{0J}_{i,jk} y^{IJ}_j y^{IJ}_k + \varepsilon_{ijk} \bigg( \varphi_j^{1J} + \varphi_{j,m}^{0J} y^{IJ}_m + \omega_{j,m}^{1J} y^{IJ}_m + \frac{1}{2} \omega^{0J}_{j,mn} y^{IJ}_m y^{IJ}_n \bigg) \bar c_{k}^{J} - \varepsilon_{ijk} \varphi_j^{1I} \bar c_{k}^{I} \bigg] e^{IJ}_{\alpha i} 
\end{split}
\end{equation}
where $\varepsilon_{ijk}$ is the Levi-Civita permutation symbol. One should consider that all length type variables have been changed into their $\mathcal{O}(\eta^{0})$ counterparts by using the rule of separation of scales, Equation \ref{scale-link-1}: $r=\eta \bar{r}$, $c_{k}^I=\eta \bar{c}_{k}^I$, $c_{k}^J=\eta \bar{c}_{k}^J$. It should be mentioned that the superscript $IJ$ is dropped whenever interchanging $I$ and $J$ does not change the sign of a quantity such as $\epsilon_{\alpha}$.
	
Having the multiple scale definition of facet strains and considering facet elastic constitutive equations $t^{(\cdot)}_{\alpha} = E_\alpha \epsilon^{(\cdot)}_\alpha$, multiple scale definition of facet elastic tractions can be written as $t_{\alpha} = \eta^{-1} t^{-1}_{\alpha} +  t^{0}_{\alpha} + \eta t^{1}_{\alpha}$. It can be proven \cite{Rezakhani-JMPS} that this asymptotic form of the facet tractions is also valid for the case of facet nonlinear constitutive equations. Using the derived asymptotic form of the facet tractions in the rescaled form of the force and moment equilibrium equations of particle $I$, Equation \ref{motion-1}, one can derive the equilibrium equations governing the problem at different scales as follows
\begin{equation} \label{motion--1-sep}
\mathcal{O}(\eta^{-2}): \hspace{0.1in} \sum_{\mathcal{F}_I}{\bar{A} t^{-1}_{\alpha} {\mathbf e}_\alpha^{IJ}} = \mathbf{0} \hspace{0.5in} \sum_{\mathcal{F}_I}{\bar{A} \left(\bar{\mathbf{c}}^I \times t^{-1}_{\alpha} {\mathbf e}_\alpha^{IJ} \right)} = \mathbf{0}
\end{equation}
\begin{equation} \label{motion-0-sep}
\mathcal{O}(\eta^{-1}): \hspace{0.1in} \sum_{\mathcal{F}_I}{\bar{A} t^{0}_{\alpha} {\mathbf e}_\alpha^{IJ}} = \mathbf{0} \hspace{0.5in} \sum_{\mathcal{F}_I}{\bar{A} \left(\bar{\mathbf{c}}^I \times t^{0}_{\alpha} {\mathbf e}_\alpha^{IJ} \right)} = \mathbf{0}
\end{equation}
\begin{equation} \label{motion-1-sep}
\mathcal{O}(\eta^{0}): \hspace{0.1in} \sum_{\mathcal{F}_I}{\bar{A}\, t^1_{\alpha} {\mathbf e}_\alpha^{IJ}} + \bar{V}^I \mathbf{b}^0 = \mathbf{0} \hspace{0.5in} \sum_{\mathcal{F}_I}{\bar{A} \left(\bar{\mathbf{c}}^I \times t^{1}_{\alpha} {\mathbf e}_\alpha^{IJ} \right)} = \mathbf{0}
\end{equation}

In the derivation of above equations, particle force and moment balance equations are rescaled through $\bar{A}=A/\eta^2$; $\bar{V}^I=V^I/\eta^3$. This is carried out in order to properly collect terms of the same order and obtain governing equations for separate scales. Full derivation of Equations \ref{motion--1-sep} through \ref{motion-1-sep} is reported in \cite{Rezakhani-JMPS}.

\noindent\textbf{RVE rigid body motion}. From the $\mathcal{O}(\eta^{-2})$ equilibrium equations, one concludes that $t_\alpha^{-1}=0$ and $\epsilon_{\alpha}^{-1}=0$, which, recalling the definition of $\epsilon_{\alpha}^{-1}=0$ in Equation \ref{eps-expansion-minus}, represents the rigid body motion of the RVE. Therefore, one can write
\begin{equation}\label{U0}
u_i^0(\mathbf{X}, \mathbf{y}) = v_i^0(\mathbf{X}) + \varepsilon_{ijk} y_k \omega_j^{0}(\mathbf{X}) 
\end{equation}

It is worth mentioning that the field variables $\mathbf{v}^0$ and $\mb \omega^{0}$, macroscopic displacement and rotation vectors, are only dependent on macroscopic coordinate system $\bf X$, which implies that these quantities varies smoothly in the macro-scale material domain, while they are constant over the RVE domain. 

\noindent\textbf{Fine-scale equations governing the RVE problem.} Equation \ref{motion-0-sep} governs the RVE problem, which reads
\begin{equation}\label{RVE-1}
\sum_{\mathcal{F}_I}{{A}\, t^{0}_{\alpha} {\boldsymbol{e}}_\alpha^{IJ}} = 0 \hspace{0.5in} \sum_{\mathcal{F}_I}{A \left(\mathbf{c}^I \times t^{0}_{\alpha} {\mathbf e}_\alpha^{IJ} \right)} = \mathbf{0}  
\end{equation}
	
Equation \ref{RVE-1} is the force and moment equilibrium equations of every single particle inside the RVE subjected to $\mathcal{O}(\eta^{0})$ facet traction $t_\alpha^0$ vector, which, in turn, is a function of $\epsilon^0_\alpha$ and can be written as \cite{Rezakhani-JMPS}
\begin{equation}\label{eps-expansion-zero'}
\epsilon_{\alpha}^0 =\bar{r}^{-1} \left( u_i^{1J} - u^{1I}_i + \varepsilon_{ijk}  \omega_j^{1J} \bar c_{k}^{J} - \varepsilon_{ijk} \omega_j^{1I} \bar c_{k}^{I}\right) {e}^{IJ}_{\alpha i} + P^\alpha_{ij} \left(\gamma_{ij}  
+ \varepsilon_{jmn}  \kappa_{im} y_{n}^{c} \right) 
\end{equation}

where $\gamma_{ij} = v^{0}_{j,i} - \varepsilon_{ijk} \omega_k^{0}$, $\kappa_{ij}=\omega_{j,i}^{0}$ are the macroscopic strain and curvature tensors, respectively. The vector $\mbf{y}^c$ is the position vector of the facet centroid shared between particles $I$ and $J$ in the local lower-scale coordinate system as shown in Figure \ref{TwoScaleAnalysis}a. $P^\alpha_{ij} = n^{IJ}_i e^{IJ}_{\alpha j}$ is an operator to project macroscopic strain and curvature tensors onto the RVE facets as normal or tangential strain components. Recalling Equation \ref{eps-1}, one can see that the first term in Equation \ref{eps-expansion-zero'} is the fine-scale definition of the facet normal and tangential strains written in terms of lower-scale displacement and rotation fields $\bold u^1$ and $\boldsymbol \omega^1$; the second term in Equation \ref{eps-expansion-zero'}, $\epsilon_{\alpha}^{c} = P^\alpha_{ij} \left(\gamma_{ij} + \varepsilon_{jmn}  \kappa_{im} y_{n}^{c} \right)$, is the projection of macroscale strain and curvature tensors on each RVE facet. In other words, to solve the RVE problem, macroscopic strain and curvature tensors should be applied on all RVE facets as negative eigenstrains, and the fine-scale solution, in terms of displacements $u_i^1$ and rotations $\omega_i^1$ of each particle, must be calculated satisfying its force and moment equilibrium equations, while periodic boundary conditions are enforced on the RVE. The solution of the equilibrium equations results in facet traction $t^0_{\alpha}$ that are then used to compute the macroscopic stress and couple tensors.
	
\noindent\textbf{Coarse-scale equations governing the macroscopic problem.} Mathematical manipulation of the equilibrium equation of $\mathcal{O}(1)$, Equation \ref{motion-1-sep}, leads to the  macroscopic translational and rotational equilibrium equations. By averaging the equilibrium equations over all RVE particles, the macro-scale translational equilibrium equation and the corresponding homogenized stress tensor are expressed as 
\begin{equation} \label{macro-eq-cont}
\sigma^0_{ji,j} + b_i = 0
\end{equation}
\begin{equation} \label{macro-stress-formula}
\sigma^0_{ij} = \frac{1}{2V_0} \sum_I \sum_{\mathcal{F}_I}{A} r t^0_\alpha P_{i j}^{\alpha}
\end{equation}
where $V_0$ is the volume of the RVE; $\rho_u=\sum_I {M}^I_u/V_0$ is the mass density of the macroscopic continuum. Equation \ref{macro-eq-cont} is the classical partial differential equation governing the equilibrium of continua whereas Equation \ref{macro-stress-formula} provides the macroscopic stress tensor though homogenizing the solution of the RVE problem. In addition, the final macro-scale rotational equilibrium equation and the corresponding macroscopic moment stress tensor are derived as
\begin{equation} \label{macro-rotational-final} 
\begin{gathered}
{\epsilon}_{ijk} {\sigma}_{ij}^0 + \frac{\partial \mu^0_{ji}}{\partial x_j} = 0
\end{gathered}
\end{equation}
\begin{equation} \label{macro-momentstress-formula} 
\begin{gathered}
\mu^0_{ij} = \frac{1}{2V_0}\sum_I \sum_{\mathcal{F}_I} {A}r t_{\alpha}^0 Q_{ij}^{\alpha}
\end{gathered}
\end{equation}
where the projection matrix $Q_{ij}^{\alpha}$ is defined as $Q_{ij}^{\alpha} = n_i^{IJ} \varepsilon_{jkl} x^C_k e_{\alpha l}^{IJ}$. $\mu^0_{ij}$ is the macroscopic moment stress tensor derived based on the results of the RVE analysis, and Equation \ref{macro-rotational-final} corresponds to the classical rotational equilibrium equation in Cosserat continuum theory. One can find the derivation details of above equations in Ref. \cite{Rezakhani-JMPS}.

\begin{figure}[t!]
\centering 
{\includegraphics[width=1\textwidth]{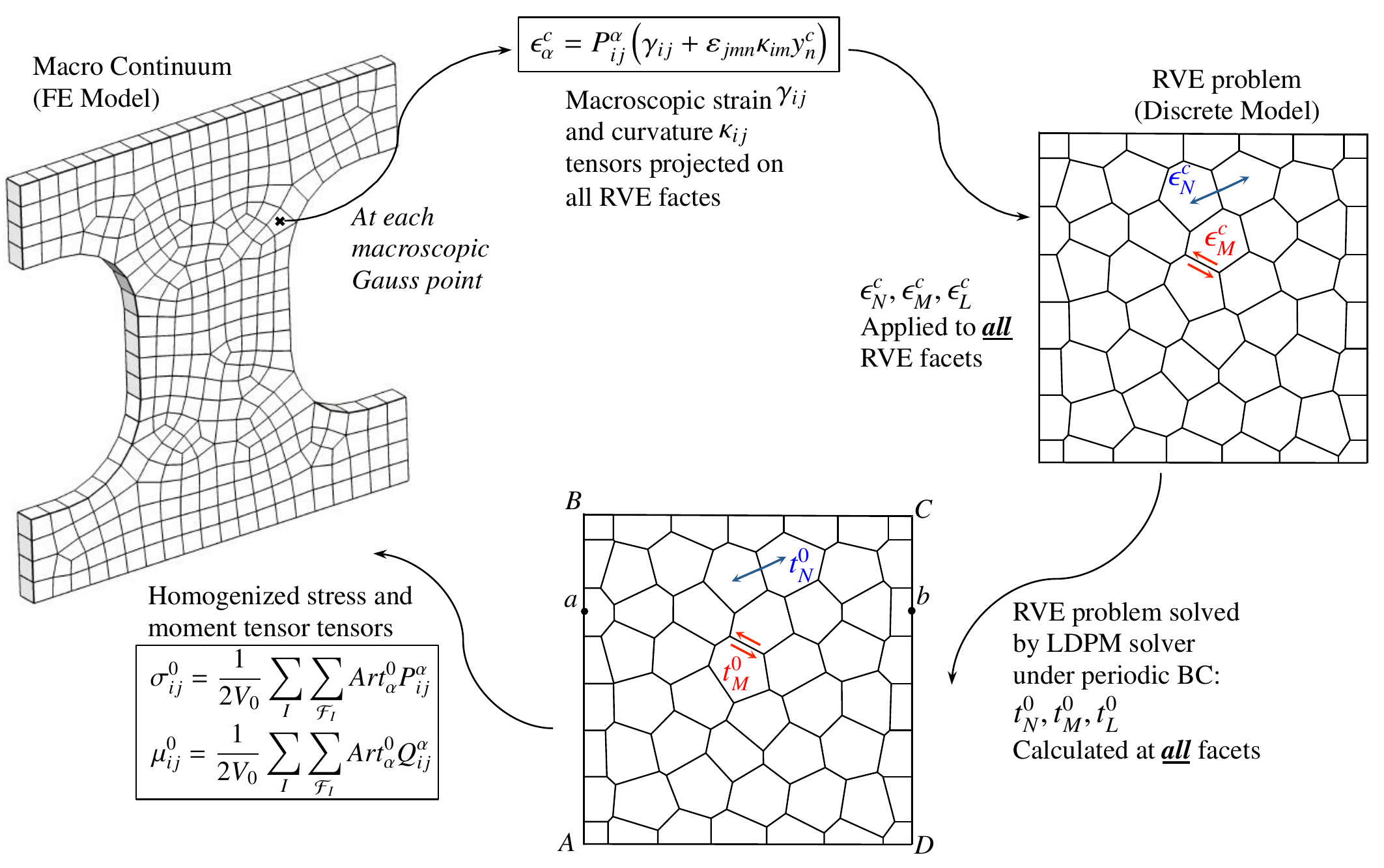}}
\caption{General framework of the classical multiscale homogenization method.}
\label{Homog-framework}
\end{figure}

\noindent\textbf{Main steps of the multiscale homogenization procedure}. The overall procedure of the discrete to continuum homogenization scheme explained in this section is illustrated in Figure \ref{Homog-framework} and can be summarized as follows:
\begin{enumerate}[label={(\arabic*)}]
\item The macro-scale material domain is discretized by FEM, and an RVE constructed by the discrete model is attached to each finite element Gauss point.
\item A global loading step is carried out, and the macroscopic strain $\gamma_{ij}$ and curvature $\kappa_{ij}$ tensors are calculated at each Gauss point of all FEs. 
\item Macroscopic strain $\gamma_{ij}$ and curvature $\kappa_{ij}$ tensors at each Gauss point are projected on all corresponding RVE facets using each facet orientation as $\epsilon_{\alpha}^{c} = P^\alpha_{ij} \left(\gamma_{ij} + \varepsilon_{jmn}  \kappa_{im} y_{n}^{c} \right)$.
\item The RVE problem is solved by applying $\epsilon_{\alpha}^{c}$ on all RVE facets and enforcing periodic boundary condition on the RVE. Fine-scale displacement $\bold u^1$ and rotation $\boldsymbol \omega^1$ of all RVE particles and tractions $t^0_\alpha$ of all RVE facet are then computed.
\item Macroscopic stress and moment stress tensors are calculated through Equations \ref{macro-stress-formula} and \ref{macro-momentstress-formula} based on the RVE response. 
\item RVE homogenized stress and moment stress tensors are then transferred back to the corresponding Gauss point and are used to update the FE nodal forces, moments, displacement, and rotations.
\item Step (1) to (6) are repeated for all macroscopic loading steps.
\end{enumerate}

As mentioned above, the RVE problem is solved under periodic boundary conditions, as shown in Figure \ref{Homog-framework}. To be able to apply periodic deformation on the RVE, each node on RVE face must have a counterpart on the other parallel face, e.g. node $a$ on edge $AB$ and node $b$ on edge $CD$, depicted in Figure \ref{Homog-framework}. These nodes are then enforced to translate and rotate equally by applying a  master-slave constraint. One should consider that for a 3D RVE, each node on an RVE edge has three counterparts on the other three parallel edges that are tied all together to deform periodically.

Several RVE and two-scale homogenization analyses presented in previous work \cite{Rezakhani-JMPS} show that, when the Lattice Discrete Particle Model (LDPM) is used as fine-scale model, the anti-symmetric part of the stress tensor is negligible, and the stress tensor is symmetric. As a result, the moment stress tensor can be assumed to be zero. Therefore, in this paper, strain and stress tensors are assumed to be symmetric, and the curvature and moment stress tensors are omitted from the calculations. This allows the use of classical FEs without rotational DOFs at the coarse-scale.

\subsection{Homogenized elastic and nonlinear RVE behavior}\label{RVE-analysis}
In this section, RVE homogenized response is investigated in the elastic and nonlinear regimes.
\subsubsection{RVE elastic analysis}\label{RVE-elastic}
Eight RVE sizes $D$ = 15, 20, 25, 35, 50, 75, 100, and 150 mm are considered to study the effect of RVE size on the homogenized elastic properties. In addition, for each RVE size, seven different particle distributions inside the RVE are analyzed to examine the effect of RVE mesh realization. This analysis is presented in \cite{Rezakhani-JMPS} to study the coefficients in elastic constitutive equations for Cosserat continua including anti-symmetric part of the stress tensor as well as the moment stress tensor as a function of curvature tensor. However, in this paper, we only investigate the effective Young's modulus and Poisson's ratio of each RVE by considering Cauchy continua as the homogeneous macroscopic material domain. This analysis is imperative for the adaptive homogenization scheme, in which RVE homogenized $E$ and $\nu$ are assigned to all macroscopic FEs at the beginning of the analysis.

To generate LDPM RVE geometries, the following parameters are considered: Minimum and maximum spherical aggregate sizes are $d_0=$ 4 mm and $d_a=$ 8 mm, respectively; cement content c = 612 $\text{kg/m}^\text{3}$; water to cement ratio w/c = 0.4; aggregate to cement ratio a/c = 2.4; Fuller curve coefficient $n_f$ = 0.42. In addition, the following parameters are used in the LDPM facet constitutive equations: $E_N = 60$ GPa, $\sigma_t = 3.45$ MPa, $\sigma_{c0} = 150$ MPa, $\alpha = 0.25$, $n_t=0.4$, $l_t=500$ mm, $r_{st}=2.6$, $H_{c0}/E_0 = 0.4$, $\mu_0 = 0.4$, $\mu_\infty = 0$, $k_{c1} = 1$, $k_{c2}=5$, $\sigma_{N0} = 600$ MPa, $\alpha=E_T/E_N=0.25$.
\begin{figure}[t!]
        \centering
        \includegraphics[width=0.7\textwidth]{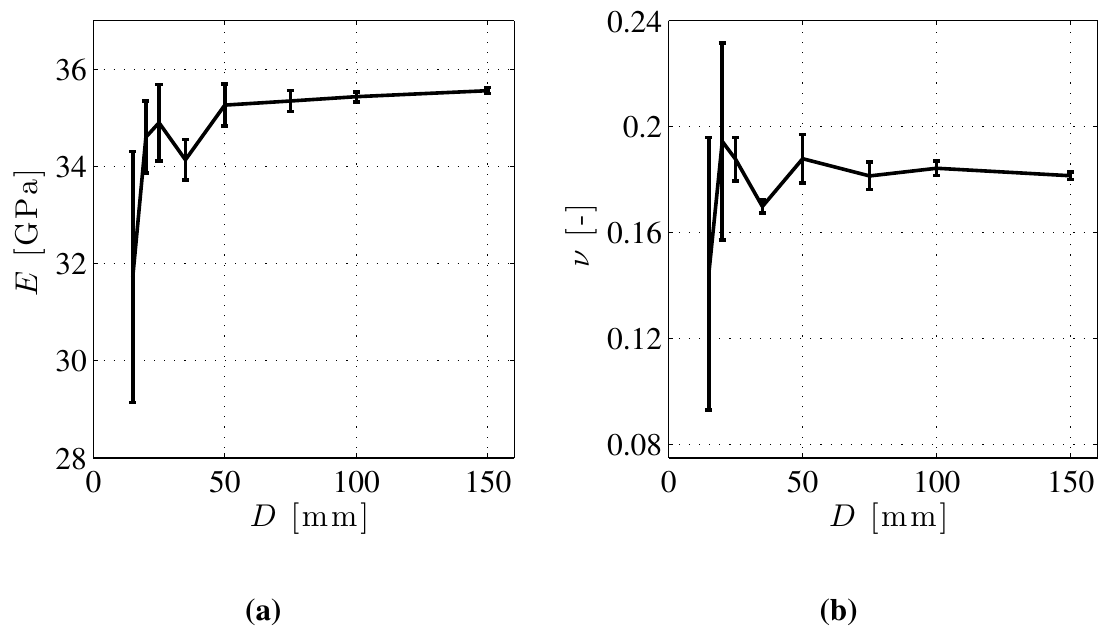}
        \caption{Variation of the effective (a) Young's modulus and (b) Poisson's ratio with respect to the RVE size.}
        \label{E-nu}
\end{figure}

Classical linear elastic constitutive equation is considered at the macroscale
\begin{equation}\label{Cauchy-Constitutive} 
\sigma_{ij} = 2\mu \gamma_{ij}^s  + \lambda \gamma_{mm} I_{ij} 
\end{equation}  
where $I_{ij}$ is the identity tensor; $\gamma_{ij}^s$ is the symmetric strain tensor; $\mu$ and $\lambda$ are Lame constants, which are related to engineering constants, Young's modulus $E$ and Poisson's ratio $\nu$, through $E = \frac{\mu (3\lambda + 2\mu)}{\lambda + \mu}$ and $\nu = \frac{\lambda}{2(\lambda + \mu)}$. These constants can be easily determined by applying proper strain tensors on the RVEs. Variation of the effective Young's modulus and Poisson's ratio with respect to the RVE size is presented in Figure \ref{E-nu}. It is shown that the homogenized values of $E$ and $\nu$ converge to a plateau as the size of the RVE is increased. Regarding these curves, one can see that a 25 mm ($D/d_a \approx 3$) RVE is an appropriate representative of the underlying material, and there is no need to further increase the RVE size. The error bars that are shown on these curves indicate the standard deviation of the results due to different RVE mesh realizations. Results show that the homogenized elastic properties become independent of the particle distribution inside the RVE, as $D/d_a$ increases. 

\begin{figure}[t]
    \centering 
	\includegraphics[width=0.8\textwidth]{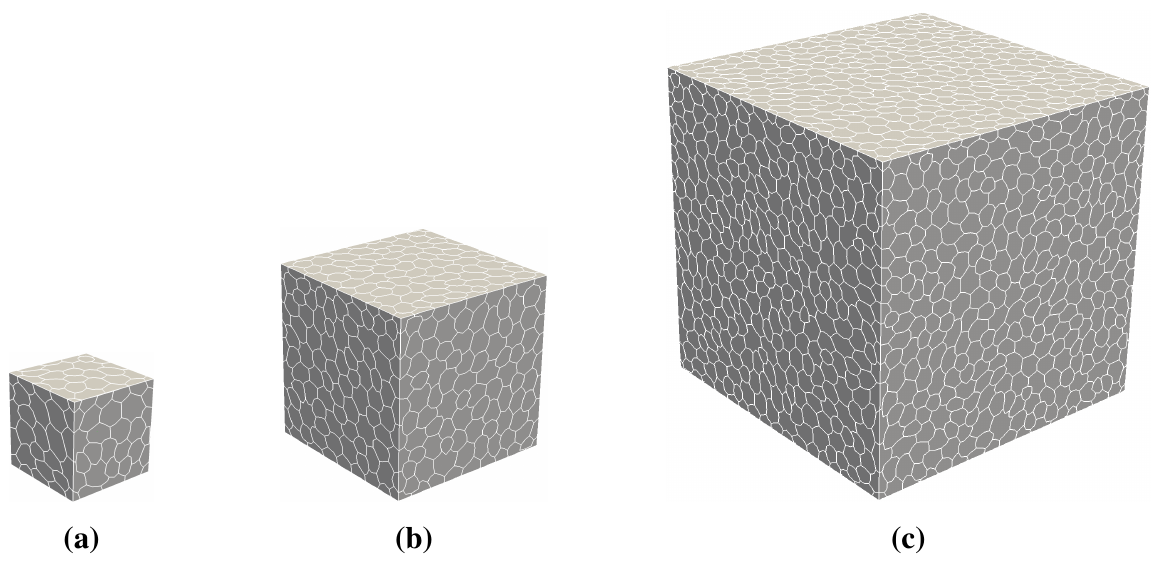}
    \caption{RVE geometry and polyhedral particle distribution: (a) 25 mm (b) 50 mm (c) 100 mm}
	\label{RVEgeom}
\end{figure}

\subsubsection{RVE nonlinear analysis}\label{RVE-nonlinear}
Nonlinear effective response of RVEs of $D$ = 25, 50, and 100 mm with six different mesh realizations for each size is studied under uniaxial tension, hydrostatic, confined (uniaxial strain), and unconfined compression while periodic boundary conditions are enforced. Each loading case is carried out on a single hexahedral FE with one integration point coupled with a material RVE. Typical particle representation and geometry of each RVE size are depicted in Figure \ref{RVEgeom}. Concrete mix design and LDPM parameters that are considered are the same as the ones that are presented in Section \ref{RVE-elastic}. 
\begin{figure}[t!]
    \centering
	\includegraphics[width=\textwidth]{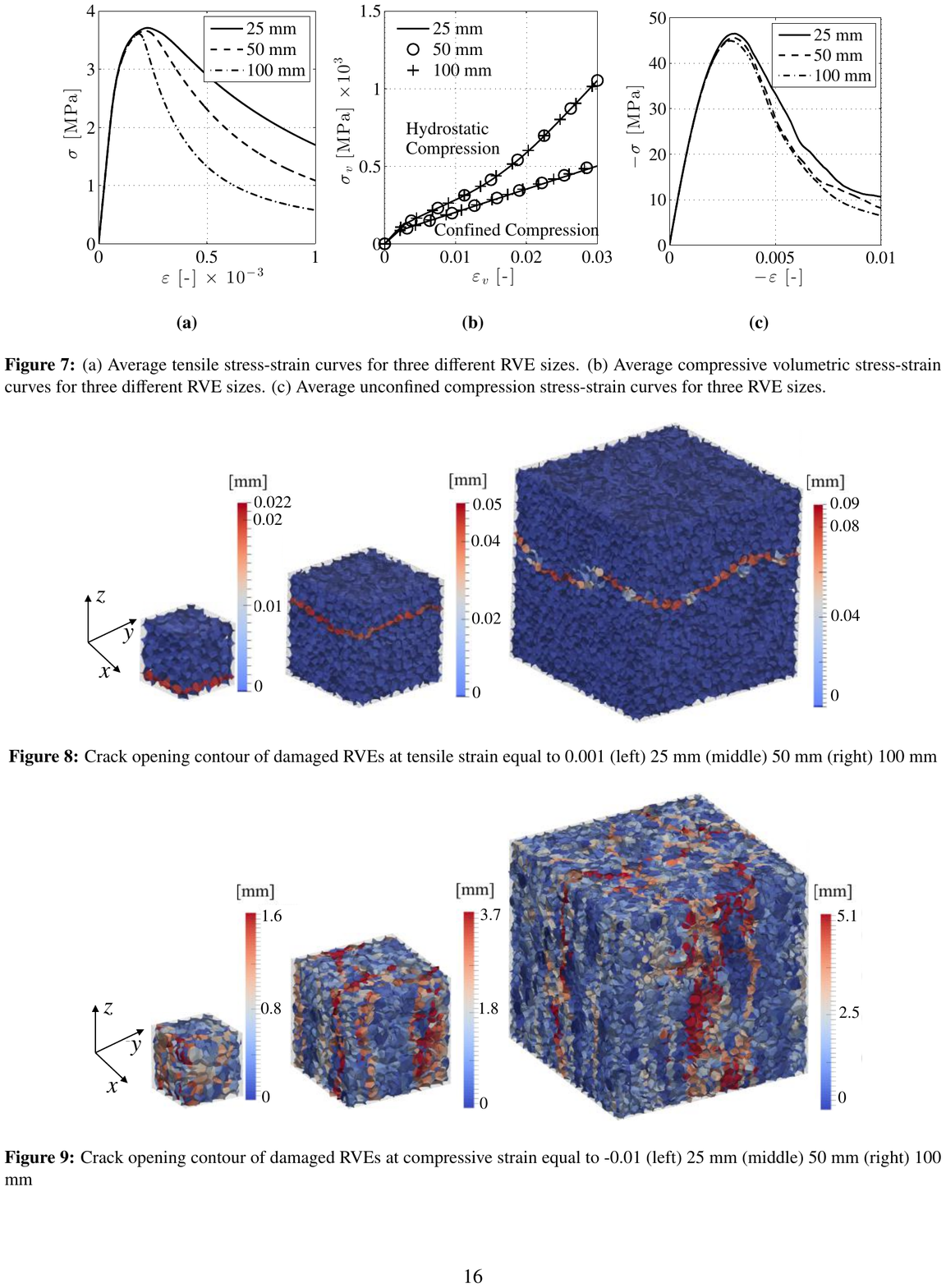}
    \caption{(a) Average tensile stress-strain curves for three different RVE sizes. (b) Average compressive volumetric stress-strain curves for three different RVE sizes. (c) Average unconfined compression stress-strain curves for three RVE sizes.}
    \label{AverageSigma}
\end{figure}

\begin{figure}[t]
\centering 
{\includegraphics[width=0.84\textwidth]{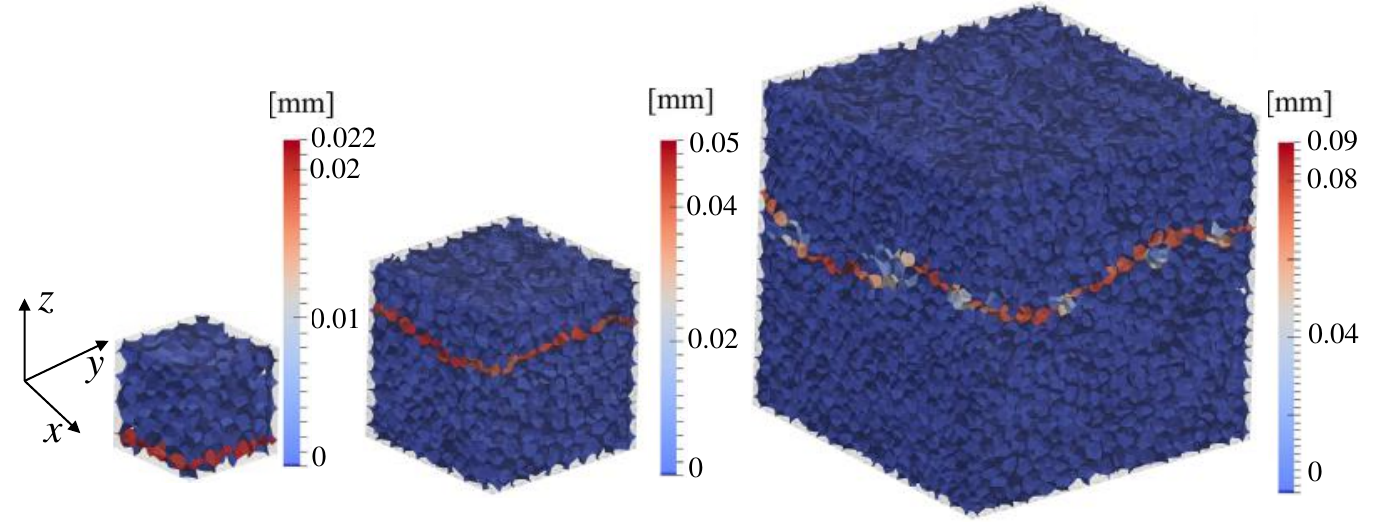}} 
\caption{Crack opening contour of damaged RVEs at tensile strain equal to 0.001 (left) 25 mm (middle) 50 mm (right) 100 mm}
\label{rve-damaged-tens}
\end{figure}

\begin{figure}[t!]
\centering 
{\includegraphics[width=0.83\textwidth]{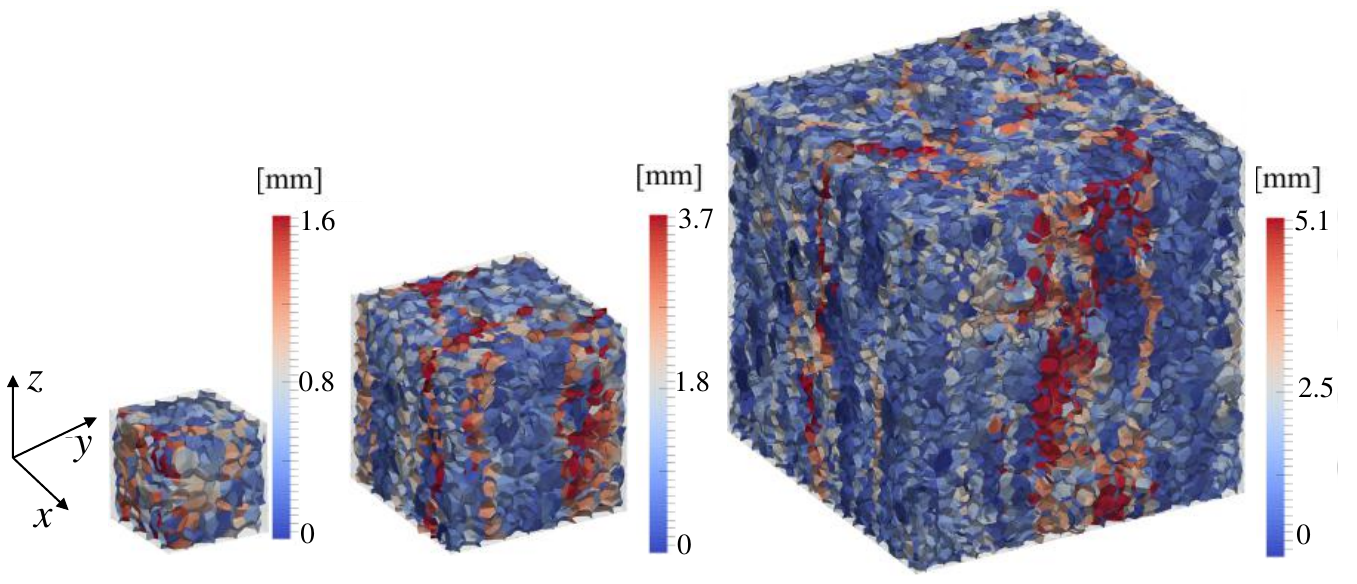}} 
\caption{Crack opening contour of damaged RVEs at compressive strain equal to -0.01 (left) 25 mm (middle) 50 mm (right) 100 mm}
\label{rve-damaged-UCC}
\end{figure}

Stress-strain results obtained from analyzing RVEs with different mesh realizations are averaged for each RVE size and presented in Figure \ref{AverageSigma}. Figure \ref{AverageSigma}a shows the homogenized nonlinear RVE behavior under tension, in which it is clearly seen that the post-peak response becomes more brittle as the RVE size is increased. Meso-scale crack opening contour of the damaged RVEs at the end of the analysis are shown in Figure \ref{rve-damaged-tens} for different RVE sizes. One can see that damage is localized in a narrow planar region of the specimen, which is the source of the size effect observed in Figure \ref{AverageSigma}a. Effective RVE response under hydrostatic and confined compressions are presented in Figure \ref{AverageSigma}b. The observed strain-hardening trend is due to the confinement effect, which leads to diffused micro-cracks over RVE volume. On the other hand, RVE softening response under unconfined compression loading condition, as shown in Figure \ref{AverageSigma}c, is associated with the vertical splitting cracks generated due to the RVE lateral expansion. Figure \ref{rve-damaged-UCC} presents the crack opening contour of the damaged RVEs, in which one can see multiple splitting cracks developed over the RVE volume at the end of the loading process. It must be noted that the RVE effective response under hydrostatic, confined, and unconfined compression is not dependent on the RVE size, as presented in Figures \ref{AverageSigma}b and c. This is due to the fact that, on the contrary to the tensile fracture, damage is distributed throughout the specimen and does not localize.

\section{Adaptive Multiscale Homogenization}\label{adaptive-homog}
Concrete, rock, and other quasi-brittle materials are characterized by strain-softening behavior, which leads to strain localization and stress redistribution. This means that damage tends to localize in a certain region of the material domain, while the rest of the material domain unload and/or remains in elastic regime. Thus, assigning material RVE to all finite elements that are used to discretize the macroscopic material domain is unnecessary, and it increases the computational cost tremendously. Therefore, defining a criterion to determine which finite elements enter the nolinear regime and must be assigned with a material RVE can be highly beneficial. This can be obtained by starting the analysis by considering elastic isotropic constitutive behavior for all FEs, and without assigning RVEs to any macroscopic FE in advance. When a finite element meets a criterion, an RVE is assigned to that finite element. Next, the inserted RVE is loaded to the level of the finite element strain tensor and is used as the element constitutive equation for the rest of the analysis. In the next section, an appropriate criterion is developed, then some numerical examples are solved to investigate the efficiency of the proposed adaptive homogenization framework. 

\subsection{RVE insertion criterion}
The Ottosen criterion, which is a combination of Rankine and Drucker-Prager criteria, is widely used to analyze concrete behavior, and it is considered in this section. The Ottosen criterion in the Haigh-Westergard space is stated as \cite{Jirasek-1}
\begin{equation} \label{ottosen}
f(\xi,\rho,\theta) = c_1 \xi + c_2 \rho r(\theta) + c_3 \rho^2 -1 = 0
\end{equation}
where $\xi$ and $\rho$ are the Haigh-Westergard space variables defined as
\begin{equation} \label{Haigh-Westergard-param}
\xi = \frac{I_1}{\sqrt{3}} = \sqrt{3} \sigma_v; \hspace{0.4 in} \rho = \sqrt{2 J_2} = \sqrt{3} \tau_{oct}
\end{equation}
where $c_1$, $c_2$, and $c_3$ are constants. $I_1 = 3\sigma_v=\sigma_{1}+\sigma_{2}+\sigma_{3}$ is the first invariant of the stress tensor in which $\sigma_v$ is the volumetric stress, and $\sigma_1$, $\sigma_2$, and $\sigma_3$ are the principal stresses. $J_2 = \frac{3}{2}\tau_{oct}^2 = \frac{1}{6}[(\sigma_1-\sigma_2)^2+(\sigma_2-\sigma_3)^2+(\sigma_1-\sigma_3)^2]$ is the second invariant of the deviatoric stress tensor, and $\tau_{oct}$ is the octahedral shear stress. $r{(\theta)}$ defines the shape of the deviatoric section:
\begin{equation}\label{rthera}
r{(\theta)} = 
\begin{dcases}
\text{cos}\bigg(\frac{1}{3} \text{arccos}(K \text{cos3}\theta)\bigg) & \text{if }  \text{cos3}\theta \geq 0\\
\text{cos}\bigg(\frac{\pi}{3} - \frac{1}{3} \text{arccos}(-K \text{cos3}\theta)\bigg) & \text{if } \text{cos3}\theta \leq 0
\end{dcases}
\end{equation}
in which $\theta$ is the Lode angle defined as
\begin{equation} \label{theta}
\text{cos3}\theta = \frac{3\sqrt{3}}{2} \frac{J_3}{J_2^{3/2}}
\end{equation}
and $J_3 = (\sigma_1-\sigma_v)(\sigma_2-\sigma_v)(\sigma_3-\sigma_v)$ is the third invariant of the deviatoric stress tensor. $K$ is the shape factor:
\begin{equation} \label{Kfactor}
K = 1 - 6.8\bigg(\frac{f_t}{f_c} - 0.07\bigg)^2
\end{equation}
in which $f_t$ and $f_c$ are typically the tensile and compressive strengths, respectively. To calibrate $c_1$, $c_2$, and $c_3$, three loading cases of uniaxial tension, uniaxial compression, and equi-biaxial compression are considered, and for each case Haigh-Westergard space variables $\xi$, $\rho$, and $\theta$ are calculated and presented in Table \ref{ottosen-param}. $f_b$ is the equi-biaxial compressive strength. Using these parameters and solving three equations with three unknowns, $c_1$, $c_2$, and $c_3$ can be obtained.  

\begin{table}[t!]
\centering
	\begin{tabular}{ |c || c c | c c | c c c | p{1cm}|} 
		\hline 
		stress state & $\sigma_1$ & $\sigma_1$ & $I_1$ & $I_2$ & $\xi$ & $\rho$ & $\theta$ \\ \hline
		uniaxial tension & $f_t$ & 0 &  $f_t$ & $f_t^2/3$ & $f_t$/$\sqrt{3}$  & $f_t$/$\sqrt{2/3}$ & 0 \\
		uniaxial compression & 0 &  $-f_c$ & $-f_c$ & $f_c^2/3$ & $-f_c$/$\sqrt{3}$  & $f_c$/$\sqrt{2/3}$ & $\pi$/3\\
		biaxial compression & $-f_b$ &  $-f_b$ &  $-2f_b$ & $f_b^2/3$ & $-2f_b$/$\sqrt{3}$  & $f_b$/$\sqrt{2/3}$ & 0\\  \hline
	\end{tabular} 
	\caption{Principal stresses, stress invariants, and Haigh-Westergard space variables at failure \cite{Jirasek-1}.}
	\label{ottosen-param}
\end{table}

\begin{figure}[b!]
\centering 
{\includegraphics[width=0.7\textwidth]{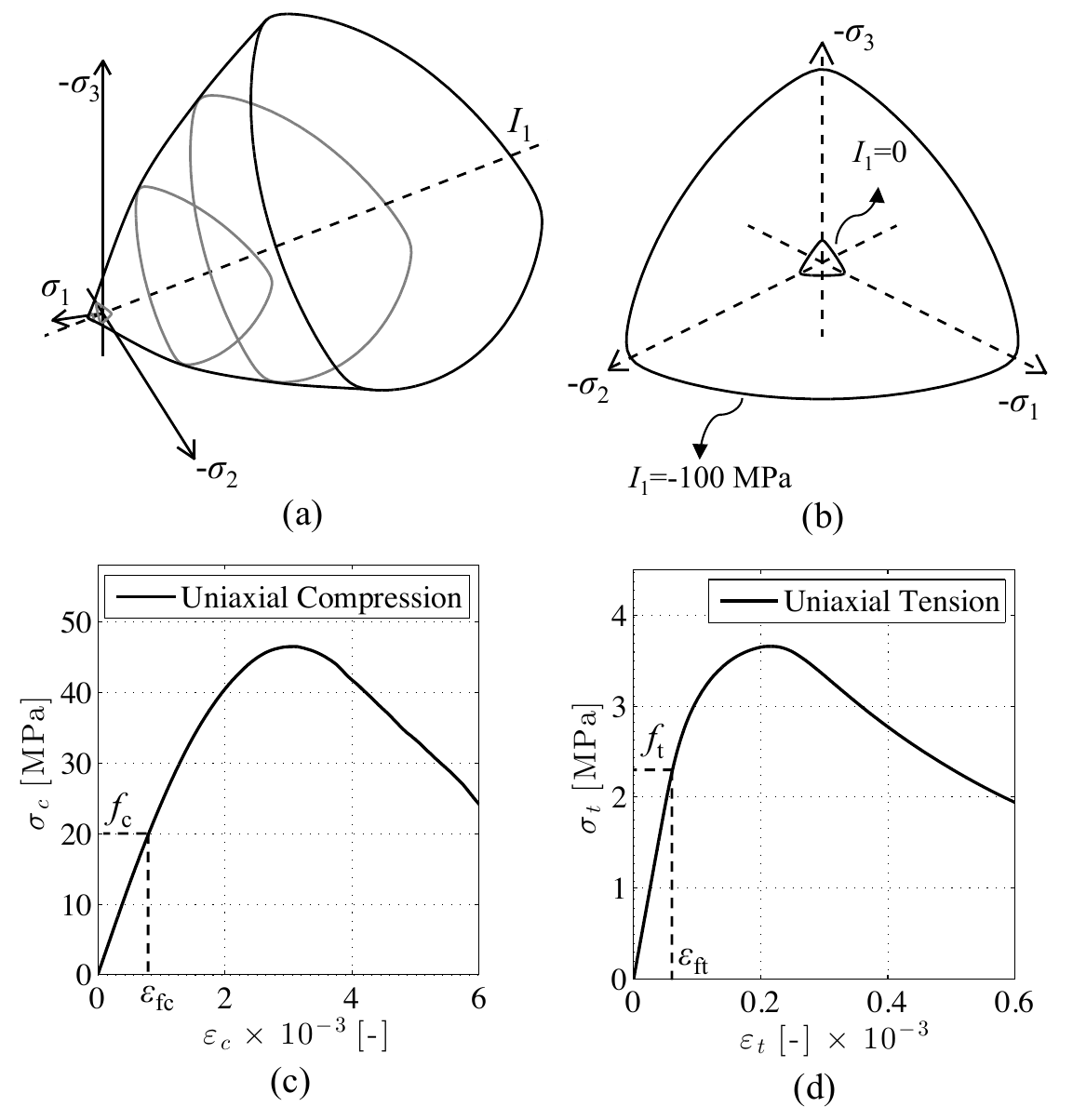}}
\caption{(a) Ottosen failure envelope. (b) Failure envelope for hydrostatic stresses equal to 0 and -100 MPa. RVE behavior under uniaxial (c) tension and (d) compression. }
\label{Ottosen}
\end{figure}

\begin{figure}[t!]
\centering 
{\includegraphics[width=\textwidth]{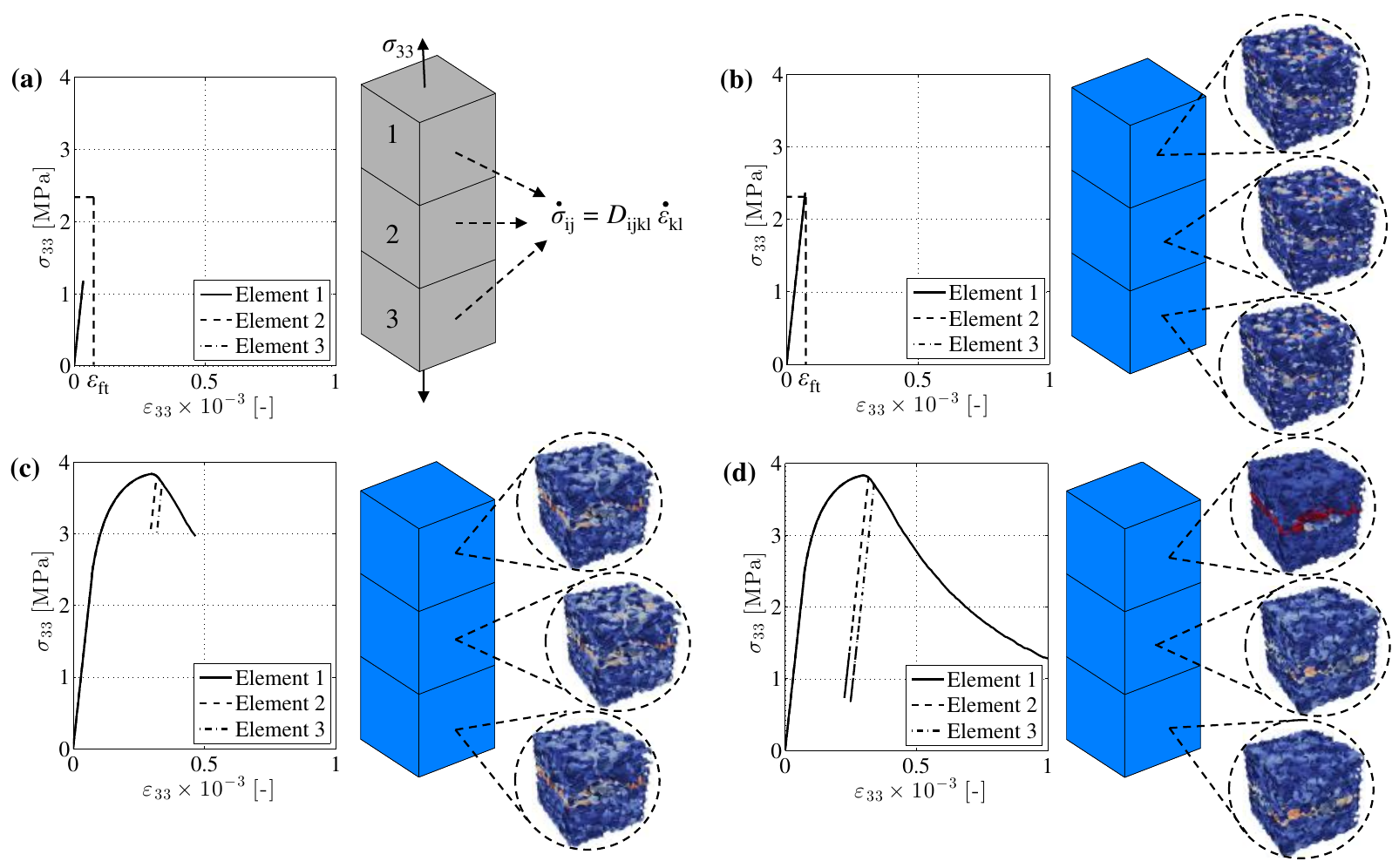}}
\caption{Adaptive multiscale homogenization general procedure. (a) Elastic constitutive equations used for all elements. (b) All elements satisfy the criterion, and a RVE is assigned to each element. (c) Two elements unload, and one enters the softening branch. (d) Strain localization in the RVE assigned to element 1.}
\label{3Elem}
\end{figure}

A generic RVE response under uniaxial tension and compression is depicted in Figures \ref{Ottosen}a and b. One can see that the stress strain curves consist of a linear elastic, nonlinear hardening, and softening post-peak parts. Since linear elastic  constitutive equation is assigned to all finite elements at the beginning of the analysis, the RVE insertion criteria must be calibrated for the linear elastic domain. Therefore, when the inserted RVE is loaded to the level of finite element strain tensor, the homogenized stress tensor calculated from the RVE is approximately equal to the element stress tensor obtained from elastic constitutive equations. In such a way, sudden changes of the system energy level, which might be due to the mismatch of macroscopic and homogenized stress tensors at that Gauss point, are prevented. Therefore, parameters $f_t$ and $f_c$ employed in calibrating the Ottosen criterion are set to the tensile and compressive linear elastic limit stresses, as shown in Figures \ref{Ottosen}c and d, and are calculated as 2.35 and 20 MPa, at $\varepsilon_{ft}=$6.7$\times 10^{-5}$ and $\varepsilon_{fc}=$5.7$\times 10^{-4}$, respectively. In addition, $f_b$ is approximated as $f_b = 1.1 f_c = 22.11$ Mpa. Using these parameters, one obtains $c_1 = 4.775/f_c$, $c_2 = 7.048/f_c$, $c_3 = 0.88/f^2_c$. It must be noted that $f_b$ and $f_c$ are calculated such that the stress values obtained from the RVE response at $\varepsilon_{ft}$ and $\varepsilon_{fc}$ is less than 5$\%$ different from the stress values computed by elastic constitutive equation at the same strain levels. The curves presented in Figures \ref{Ottosen}c and d are generated through nonlinear analysis of a 50 mm RVE with LDPM parameters equal to the ones presented in Section \ref{RVE-elastic}. Since changing the RVE size only affects the post peak behavior and has minor effect on the elastic response, as long as $D/d_a$ is large enough, the calibrated $c_1$, $c_2$, and $c_3$ do not change when a different RVE size with the same LDPM parameters is employed.

It should be noted that the elastic material properties assigned to the FEs at the beginning of the analysis is computed by elastic analysis of the specific RVE, which will be assigned to each FE meeting the criterion during the analysis, as explained in Section \ref{RVE-elastic}. 

\noindent\textbf{Adaptive multiscale homogenization procedure}. To clarify the computational strategy in the proposed adaptive homogenization framework, a simple concrete prism is modeled with 3 finite elements and is loaded under axial tension, as shown in Figure \ref{3Elem}. The numerical procedure can be described as follows:

\begin{enumerate}[label={(\arabic*)}]
\item The effective young modulus $E$ and Poisson's ratio $\nu$ of the chosen RVE are calculated by elastic analysis of the RVE as explained in Section \ref{RVE-elastic}.

\item The elasticity tensor $D_{ijkl}$ is built by using the calculated effective elastic properties of the RVE, and linear elastic constitutive equations $\dot{\sigma}_{ij} = D_{ijkl} \dot{\varepsilon}_{kl}$ are employed for all FEs, as shown in Figure \ref{3Elem}a. $\dot{\sigma}_{ij}$ and $\dot{\varepsilon}_{kl}$ are the rates of stress and strain tensors.

\item The elastic constitutive equation is used for each FE until the calibrated criterion in Equation \ref{ottosen} is satisfied. Once the stress tensor at the Gauss point of a finite element calculated from the elastic constitutive equation meets the criterion, the element constitutive equation is switched from elastic constitutive equation to a meso-scale material RVE, see Figure \ref{3Elem}b.

\item The FE strain tensor is applied on the assigned RVE in order to load the RVE to the current level of the FE stress tensor, and the analysis proceeds using the inserted RVEs as the finite element constitutive equation, see Figures \ref{3Elem}c and d. It is important to point out that the RVE size is such that its volume matches the one of the FE. This ensures correct energy dissipation and size effect under strain softening.
\end{enumerate}

In Figure \ref{3Elem}c, one can see that finite elements 2 and 3 unloads, while the finite element 1 enters the softening branch. Considering Figure \ref{3Elem}d, it is observed that damage localization occurs in the RVE assigned to finite element 1, while the cracks developed over the RVEs of the finite elements 2 and 3 close, as these finite elements unload. 

\section{Numerical Results}

\subsection{Four point bending test}
\begin{figure}[t!]
\centering 
{\includegraphics[width=\textwidth]{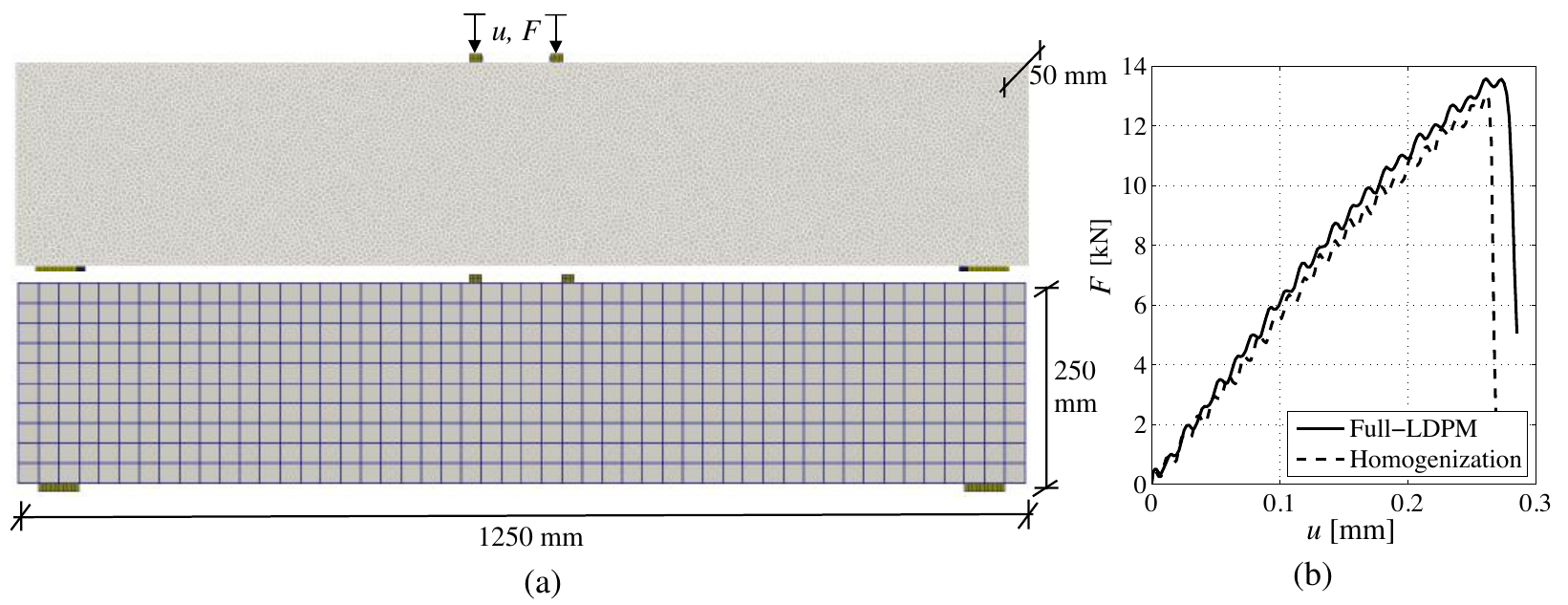}}
\caption{(a) The beam modeled by finite element and LDPM. (b) Force displacement curve obtained from homogenization and LDPM simulations.}
\label{4pbt-geom}
\end{figure}
In this section, the four point bending problem is solved by the proposed adaptive multiscale homogenization, and the results are compared to the full fine-scale analysis of the same problem. A beam of 1250 mm length, 250 mm height, and 50 mm width is selected and modeled by both FEM and LDPM, as shown Figure \ref{4pbt-geom}a. The same LDPM parameters listed in Section \ref{RVE-elastic} are used in these simulations. For the homogenization problem, solid finite elements of 25 mm size are used to discretize the macroscopic problem. In addition, a material RVE of 25 mm size is constructed to be inserted into the adaptive homogenization framework during the analysis. The finite element size and the RVE size are chosen to be equal in order to cover the same damaged area in both homogenization and full fine-scale analyses, which makes the dissipated energy to be equal in both simulations \cite{Gitman-1,Gitman-2} preserving correct energy dissipation and size effect \cite{Bazant-Book}. RVE effective elastic properties are calculated as $E = 35$ GPa and $\nu = $ 0.17 and are employed as macro-scale material properties in the adaptive framework. The beam is simply supported and is equally loaded by two loading platens located at 50 mm from midspan. The calculated force-displacement curves of the homogenization and full fine-scale analyses are plotted in Figure \ref{4pbt-geom}b. One can see that homogenization result agrees well with the fine-scale analysis in terms of both the slope of the linear part and the peak value. It is should be pointed out that the oscillations observed in the force-displacement response of the beams is due to fact that the simulations are performed with an explicit dynamic algorithm.

\begin{figure}[h!]
\centering 
{\includegraphics[width=0.95\textwidth]{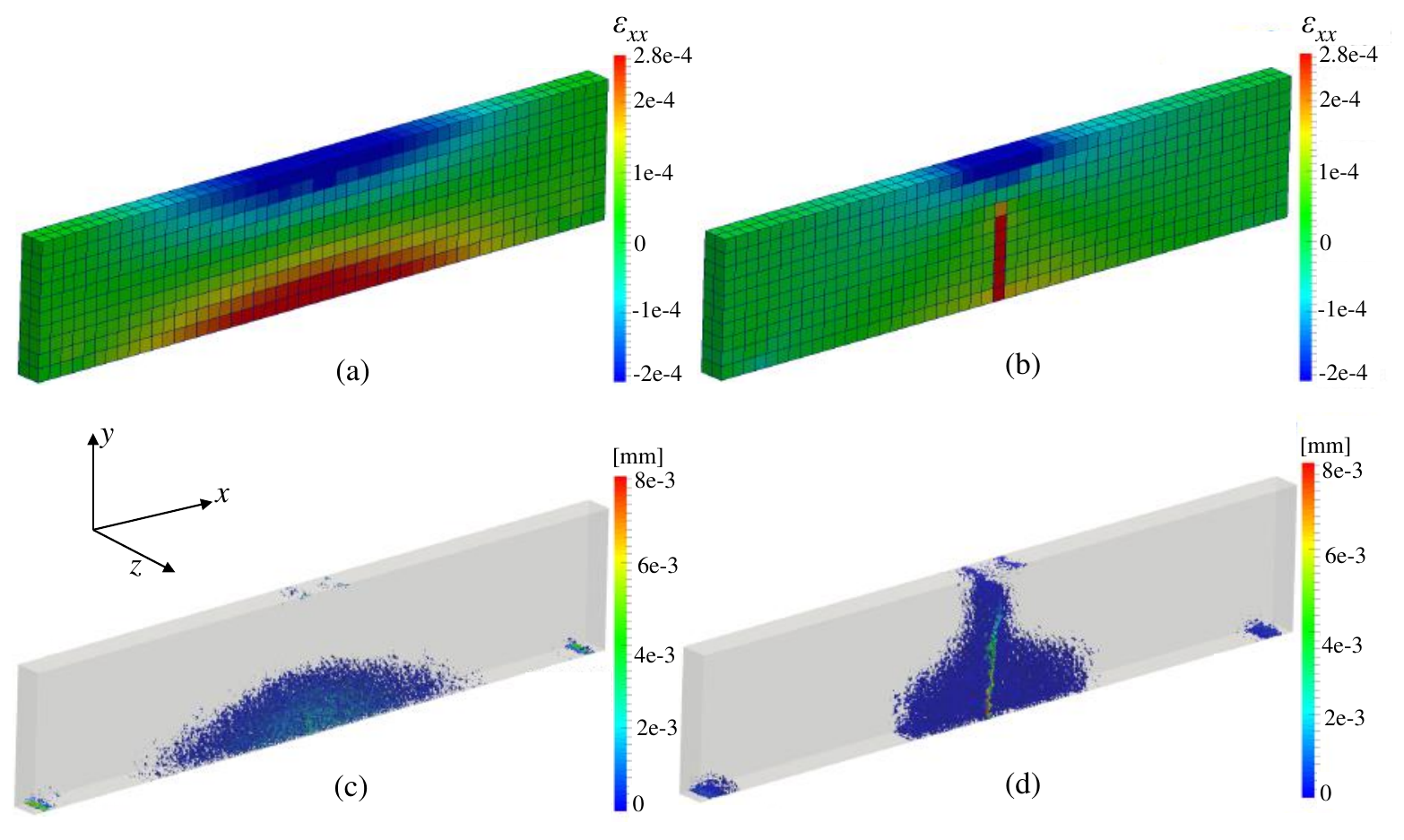}}
\caption{(a) $\varepsilon_{xx}$ contour obtained from homogenization at 0.1 mm displacement of the loading platens. (b) $\varepsilon_{xx}$ contour obtained from homogenization after the peak; (c), (d) crack opening contours from full fine-scale analysis at the same analysis instants as (a) and (b), respectively. }
\label{4pbt-damage}
\end{figure}

Figure \ref{4pbt-damage}a shows the FEs $\varepsilon_{xx}$ contour obtained from the homogenization framework at 0.1 mm displacement of the loading platens, corresponding to a pre-peak load level, see Figure \ref{4pbt-geom}b. These contour plots are comparable with the crack opening contours plotted at the same instant of the full fine-scale analysis, as shown in Figure \ref{4pbt-damage}c. One can see that the micro cracks generated in the full fine-scale analysis are diffused over the volume where positive $\varepsilon_{xx}$ is distributed in the multiscale homogenization analysis. The $\varepsilon_{xx}$ contour at failure obtained from the multiscale homogenization analysis is depicted in Figure \ref{4pbt-damage}b. The localized damage pattern is consistent with the crack opening contour of the full fine-scale analysis at failure plotted in Figure \ref{4pbt-damage}d. The homogenization framework successfully replicated the full fine-scale analysis results with respect to both the force-displacement response and the specimen damage distribution.

\begin{figure}[h!]
\centering 
{\includegraphics[width=\textwidth]{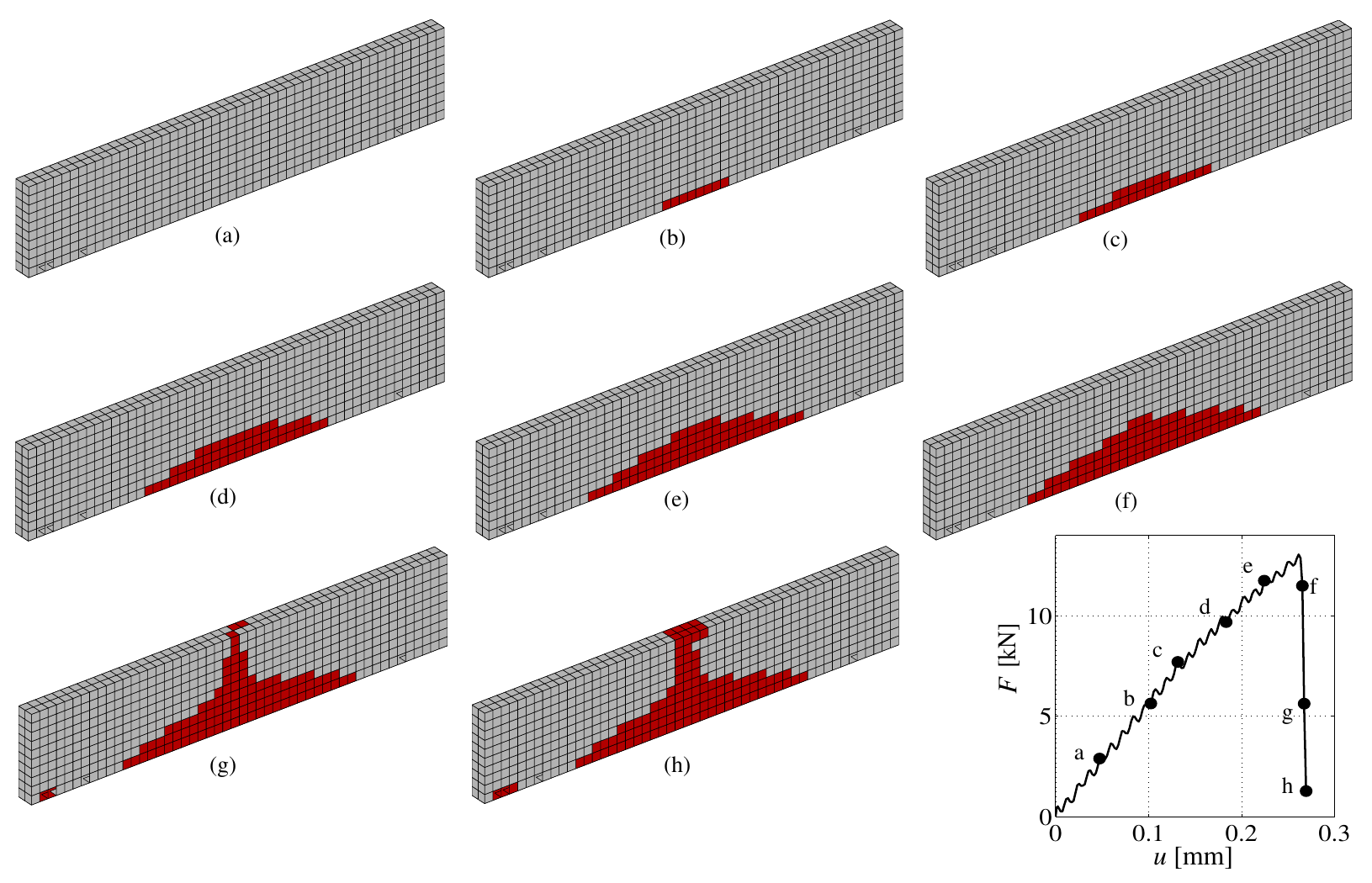}}
\caption{RVE insertion procedure at different instants of the homogenization analysis.}
\label{4pbt-RVEinsert}
\end{figure}

During the adaptive multiscale homogenization analysis, the macro-scale finite elements that satisfy the RVE insertion criterion are plotted in Figure \ref{4pbt-RVEinsert} sequentially. Each subfigure corresponds to a point on the force-displacement curve shown in Figure \ref{4pbt-RVEinsert}. One can see that up to point (a), all of the finite element are inside the linear elastic behavior envelope, and elastic constitutive equation is used for all of them. At point (b), sixteen finite elements have met the criterion, and a fine-scale material RVE is assigned to each of those elements. As the analysis continues, more finite elements meet the criterion, which can be seen from Figure \ref{4pbt-RVEinsert}c to \ref{4pbt-RVEinsert}f. It is interesting to note that up to the point (f), the shape of the finite elements stack that met the RVE insertion criterion corresponds to the distribution of the positive $\varepsilon_{xx}$ generated due to the beam bending as shown in Figure \ref{4pbt-damage}a. In addition, considering Figures \ref{4pbt-RVEinsert}g and \ref{4pbt-RVEinsert}h, the selected finite elements at points g and h, that are placed on the post-peak branch, follow the beam localized damage pattern at failure as shown in Figure \ref{4pbt-damage}b.

\subsection{L-shape specimen test}
\begin{figure}[h!]
\centering 
{\includegraphics[width=\textwidth]{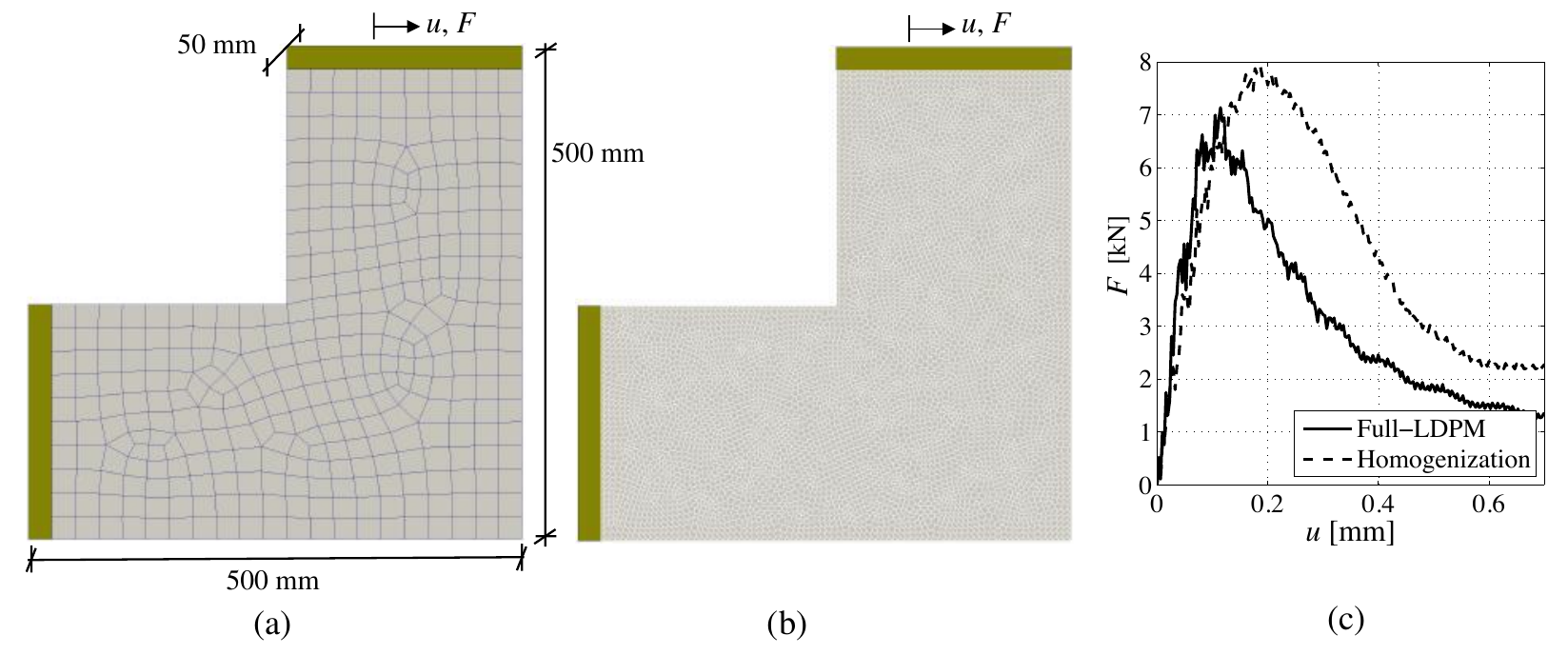}}
\caption{(a) and (b) L-shape specimen geometry modeled by finite element and LDPM, respectively. (c) Force displacement curve obtained from homogenization and LDPM simulations.}
\label{L-geom}
\end{figure}

\begin{figure}[h!]
\centering 
{\includegraphics[width=0.7\textwidth]{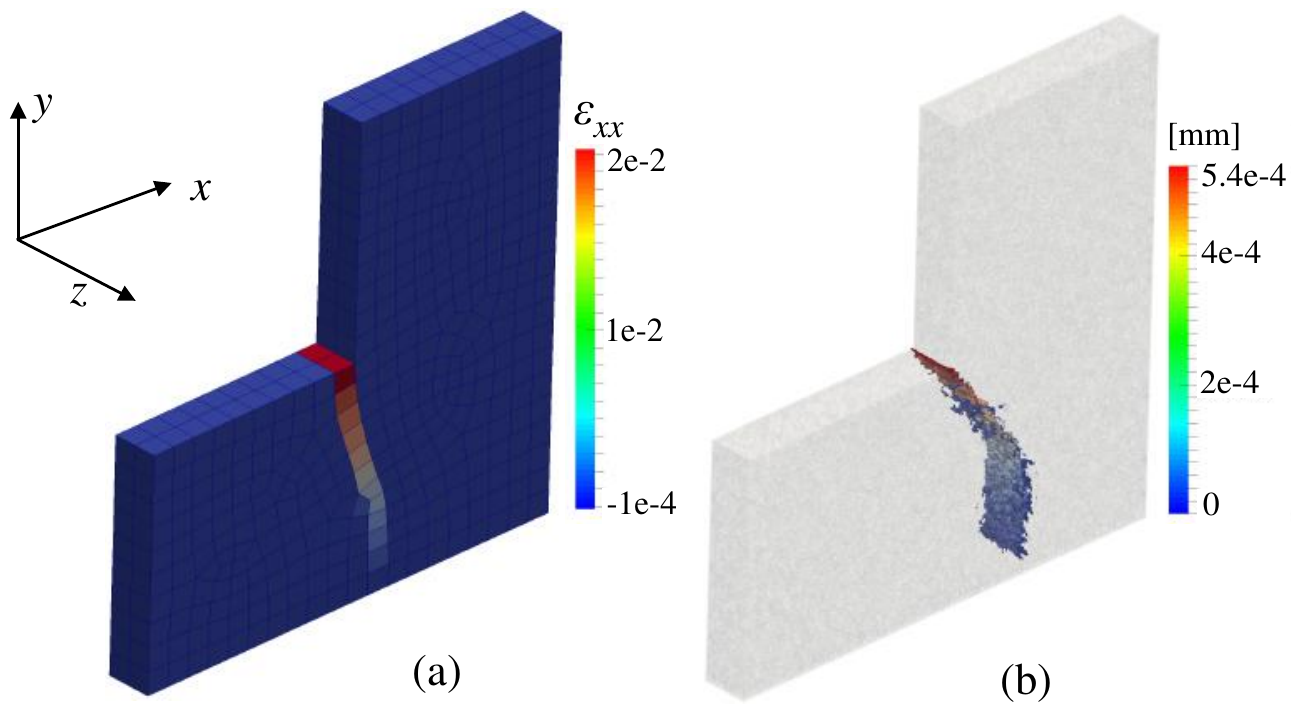}}
\caption{(a) $\varepsilon_{xx}$ contour obtained from homogenization at the end of the analysis. (c) Crack opening contour obtained from LDPM simulation at the end of the analysis.}
\label{L-damage}
\end{figure}

\begin{figure}[t!]
\centering 
{\includegraphics[width=0.8\textwidth]{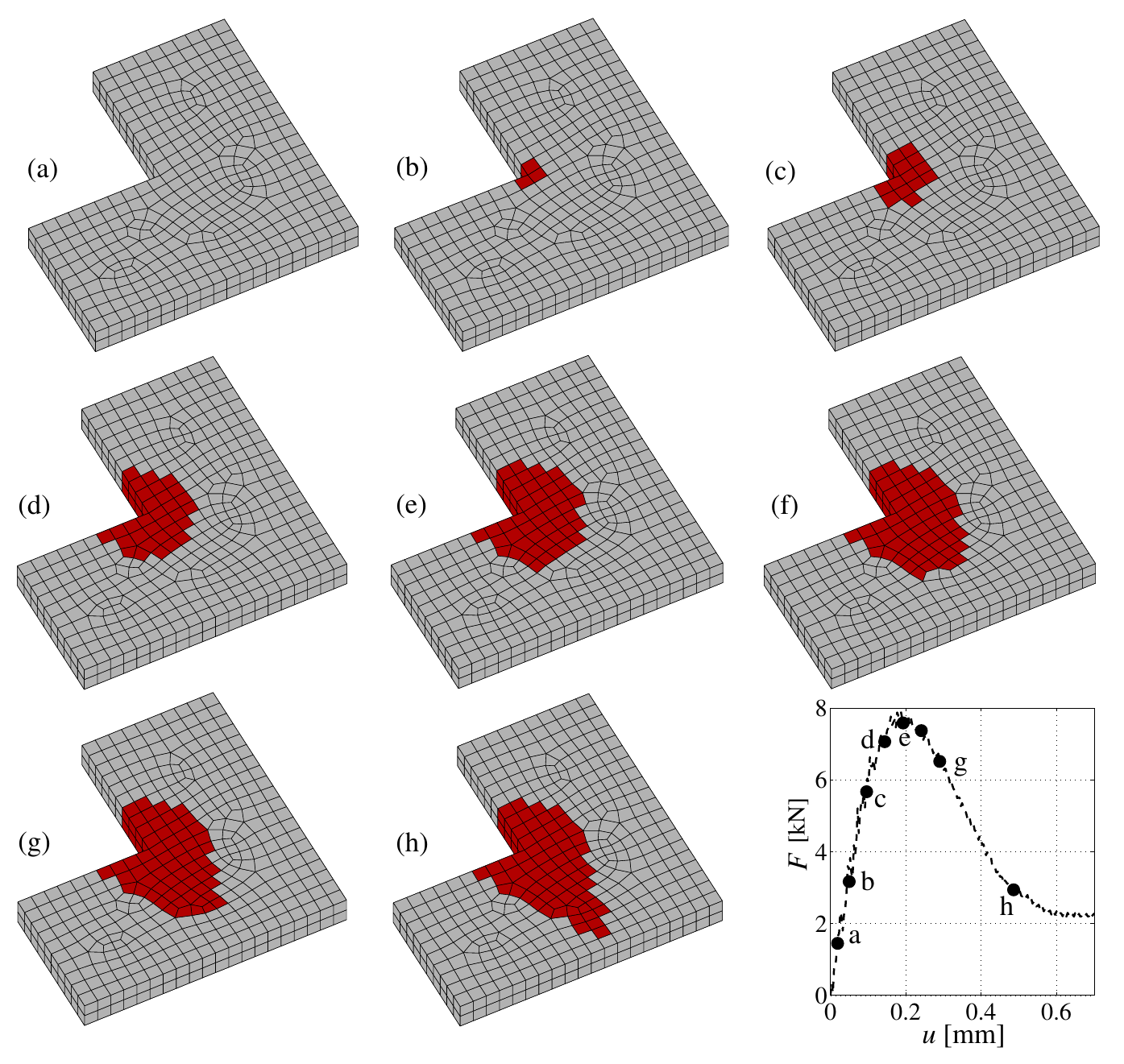}}
\caption{RVE insertion procedure at different instants of the homogenization analysis.}
\label{L-RVEinsert}
\end{figure}

An L-shape specimen subjected to a horizontal force is analyzed by both the full fine-scale particle model and the adaptive multiscale homogenization method. The generated specimens by FEM and LDPM are presented in Figures \ref{L-geom}a and b. The specimen is fixed at the left side in all directions, and is subject to a horizontal force by means of a loading plate at the top side. The same LDPM parameters listed in Section \ref{RVE-elastic} are employed in these simulations. 25 mm side hexahedral solid finite elements are used to discretize the macroscopic problem domain. Similarly to the four point bending problem, the macroscopic FEs and the constructed RVE sizes are considered to be equal, and $E = 35$ GPa and $\nu = $ 0.17 are the RVE homogenized elastic properties that are assigned to the macro-scale finite elements. Figure \ref{L-geom}c shows the force displacement curve of the loading plate obtained from the multiscale homogenization and the full fine-scale analyses. One can see that the homogenization result is in relatively good agreement with the full fine-scale analysis result with respect to both the slope of the linear part and the trend of the softening branch. There is approximately a 10$\%$ difference in the force peak value, which is due to the coarse finite element discretization of the macroscopic domain in the multiscale homogenization analysis. The damage localization in the finite element analysis is constrained to follow the possible paths provided by the finite element discretization. In the current problem, due to the coarse and roughly structured finite element mesh, possible paths that are available for the strain localization band to develop are limited, which definitely influences the results. In the classical FE simulation of such problems, the FE mesh can be refined to capture the correct localization path. However, in the homogenization analysis, the size of finite element mesh is constrained by the RVE size as its lower limit, which, as discussed in Section \ref{RVE-elastic}, must be larger than $\approx 3d_a$ to properly represent the behavior of underlying material. Therefore, numerical problems including sharp corners and inclined localization paths can be mentioned as a limitation of the homogenization framework.

The developed damage pattern in the multiscale homogenization and the full fine-scale analyses are presented in Figure \ref{L-damage}. One can see that the strain localization band obtained from the homogenization analysis approximately matches the crack opening contour captured by the full fine-scale simulation. The RVE insertion sequence is plotted in Figure \ref{L-RVEinsert}. One can see that up to point (a) shown on the force-displacement curve, no RVE is assigned to any finite element, and the tensorial elastic constitutive equation is used for all elements. At point (b), six finite elements meet the criteria, and their constitutive equations are switched from the elastic tensorial one to an RVE assigned to their Gauss points. The RVE insertion sequence begins from the sharp corner of the specimen where stress concentration occurs. Strain localization band initiates from this corner and propagates through the specimen. As the loading process continues, more finite elements are equipped with a RVE, and this can be observed as the red area in Figure \ref{L-RVEinsert} spreads. In Figure \ref{L-RVEinsert}h, one can see that the developed red area at the end of the simulation encompasses the band of finite elements where damage localization takes place as shown in Figure \ref{L-damage}a.

\begin{figure}[t!]
	\centering
	\includegraphics[width=0.7\textwidth]{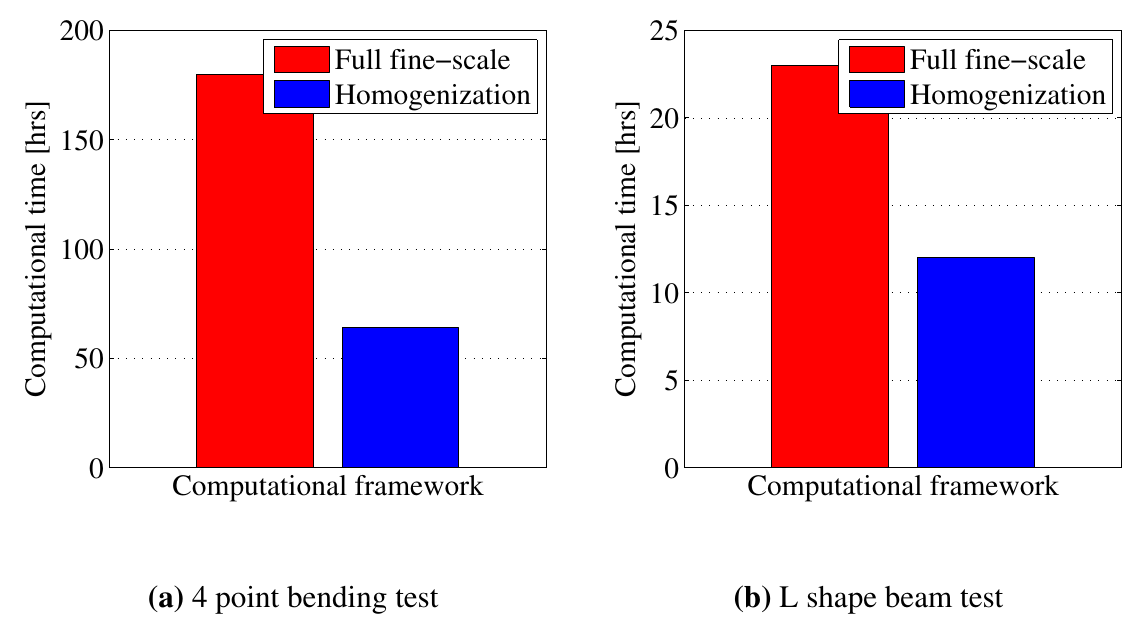}
	\caption{Computational cost comparison of the full fine-scale and the multiscale adaptive homogenization analyses.}
	\label{CompTime-Compare}
\end{figure}

\subsection{Computational cost saving}\label{comp-cost}
The most significant feature of the proposed adaptive multiscale homogenization scheme is the fact that the fine-scale material RVEs are not assigned to all macroscopic finite elements from the beginning of the analysis, as opposed to the classical homogenization method. In addition, users are not required to specify the finite elements that must be assigned with a material RVE in advance. The computational framework methodically detects the critical finite elements that must be inserted into the multiscale analysis. In the four point bending test problem, 206 finite elements out of the total 1,000 finite elements were enriched with the RVEs ($\approx 20\%$). Out of the total 680 finite elements in the L-shape specimen problem, only 132 elements met the linear elastic limit criterion ($\approx 19\%$). In both cases, the proposed adaptive scheme led to a significant computational cost saving. In Figure \ref{CompTime-Compare}, the total computational time of the full fine-scale simulations of the beam four point bending test and the L-shape specimen are reported as 180 hours and 23 hours on a single processor, respectively. With the same computational resources, it takes 64 hours (35$\%$ of the full fine-scale computational time) for the multiscale analysis of the four point bending test, and 12 hours (35$\%$ of the full fine-scale computational time) for the L-shape specimen test. 

Another source of computational time saving is the initialization time. The full fine-scale beam generated by LDPM consists of 48,550 nodes and 240,740 tetrahedral elements, while the RVE used in the multiscale analysis includes 163 nodes and 529 LDPM tetrahedral elements. Similarly, the L-shape specimen consists of 31,165 nodes and 150,767 tetrahedral elements, while the fine-scale material RVE is characterized by 154 nodes and 498 tetrahedral elements. Therefore, the amount of time necessary to generate the RVE mesh is negligible compared to the initialization time needed to generate the geometry of the full fine-scale specimens. 
 
\section{Conclusion} 
In this paper a recently published multiscale homogenization method, which couples a fine-scale discrete model to a macro-scale continuum model, is implemented in an adaptive framework. An RVE insertion criterion is defined based on the Ottosen failure criteria widely used for the simulation of concrete behavior. On the contrary to the classical homogenization method, there is no need to assign any RVE to any specific part of the macro-scale problem domain ahead of the analysis. The adaptive strategy automatically detects the finite elements that should be inserted into the homogenization framework. Therefore, computational cost is considerably decreased with respect to the classical homogenization method due to less number of RVEs that are assigned to macroscopic finite elements. The adaptive homogenization scheme provides accurate results compared to the full fine-scale simulations. 
\newline \\
\noindent\textbf{ACKNOWLEDGMENTS} \\
This material is based upon work supported by the National Science Foundation under grant no. CMMI-1435923.

\newpage
\appendix
\numberwithin{equation}{section}

\end{document}